\documentclass[sn-mathphys-num]{sn-jnl}

\usepackage{graphicx}%
\usepackage{multirow}%
\usepackage{amsmath,amssymb,amsfonts}%
\usepackage{amsthm}%
\usepackage{mathrsfs}%
\usepackage[title]{appendix}%
\usepackage{xcolor}%
\usepackage{textcomp}%
\usepackage{manyfoot}%
\usepackage{booktabs}%
\usepackage{algorithm}%
\usepackage{algorithmicx}%
\usepackage{algpseudocode}%
\usepackage{listings}%

\theoremstyle{thmstyleone}%
\theoremstyle{thmstyletwo}%

\theoremstyle{thmstylethree}%

\usepackage{amsthm}
\usepackage{pdfpages}
\usepackage{stmaryrd}
\usepackage{enumitem}
\usepackage{comment}
\usepackage{times}

\usepackage{graphicx,booktabs,listings,caption,subcaption,tikz,amsmath,amssymb,placeins,subfiles,tabto,scalerel,stackengine}
\usepackage{xcolor,colortbl}
\usepackage{multirow,float}

\usepackage{mathtools,centernot}
\usepackage{tikz, trimclip}

\usepackage{makecell}
\usepackage{cleveref}
\usepackage{pifont}

\newcommand{\tent}[1]{#1}

\definecolor{azure}{rgb}{0.0, 0.5, 1.0}
\definecolor{cadmiumorange}{rgb}{0.93, 0.53, 0.18}
\definecolor{cadmiumgreen}{rgb}{0.0, 0.42, 0.24}
\definecolor{cadmiumred}{rgb}{0.89, 0.0, 0.13}
\definecolor{cadmiumyellow}{rgb}{1.0, 0.96, 0.0}
\definecolor{applegreen}{rgb}{0.55, 0.71, 0.0}
\definecolor{greyblue}{rgb}{0.25, 0.32, 0.61}

\definecolor{otr}{rgb}{0.9725490196078431, 0.807843137254902, 0.8}

\definecolor{r}{rgb}{0.9568627450980393, 0.3686274509803922, 0.34509803921568627}
\definecolor{g}{rgb}{0.0, 0.5803921568627451, 0.2980392156862745}
\definecolor{p}{rgb}{0.5764705882352941, 0.47058823529411764, 0.8901960784313725}
\definecolor{y}{rgb}{0.9529411764705882, 0.7294117647058823, 0.2196078431372549}
\definecolor{o}{rgb}{0.9568627450980393, 0.4980392156862745, 0.2980392156862745}
\definecolor{pi}{rgb}{0.9803921568627451, 0.4980392156862745, 0.6941176470588235}

\newcommand{\mode}       {\mu}
\newcommand{\Time}       {\mathcal{T}}   
\newcommand{\Real}       {\mathbb{R}}    
\newcommand{\barp}       {\bar{p}}      

\newcommand{\nunets}     {$\nu$-nets}

\newcommand{\al}         {\gamma}  
\newcommand{\relal}      {\tilde\gamma} 

\newcommand{\Lissue}[2] {$\text{RI}_\text{#1}^\text{#2}$}

\newcommand{\proj}              {{\upharpoonright}}

\DeclareMathOperator{\gt}{gt}

\newcommand{\set}[1]{\{ #1 \}}

\newcommand{\mset}[1]{[ #1 ]}

\newcommand{\wc}[1]                {{\langle}#1{\rangle}}

\newcommand{\pre}[1]             {{}^{\bullet}{#1}}
\newcommand{\post}[1]            {#1^\bullet}

\newcommand{\Ie}{I.e., }
\newcommand{\ie}{i.e., }
\newcommand{\Eg}{E.g., }
\newcommand{\eg}{e.g., }

\newcommand{\cf}{c.f., }

\begin{document}

\title[Article Title]{Article Title}


\author*[1]{\fnm{Dominique} \sur{Sommers}}\email{d.sommers@tue.nl}

\author[1]{\fnm{Natalia} \sur{Sidorova}}\email{n.sidorova@tue.nl}

\author[1]{\fnm{Boudewijn} \spfx{van} \sur{Dongen}}\email{b.f.v.dongen@tue.nl}

\affil[1]{\orgdiv{Mathematics and Computer Science}, \orgname{Eindhoven University of Technology}, \orgaddress{\city{Eindhoven}, \country{the Netherlands}}}

\title{
A Ground Truth Approach for Assessing Process Mining Techniques
}

\abstract{
The assessment of process mining techniques using real-life data is often compromised by the lack of ground truth knowledge, the presence of non-essential outliers in system behavior and recording errors in event logs. Using synthetically generated data could leverage ground truth for better evaluation. Existing log generation tools inject noise directly into the logs, which does not capture many typical behavioral deviations. Furthermore, the link between the model and the log, which is needed for later assessment, becomes lost.

We propose a ground-truth approach for generating process data from either existing or synthetic initial process models, whether automatically generated or hand-made. This approach incorporates patterns of behavioral deviations and recording errors to produce a synthetic yet realistic deviating model and imperfect event log.
These, together with the initial model, are required to assess process mining techniques based on ground truth knowledge. 
We demonstrate this approach \tent{to create datasets of synthetic process data for three processes, one of which we used in
}a conformance checking use case, focusing on \tent{the assessment of} (relaxed) systemic alignments to expose and explain deviations in modeled and recorded behavior. Our results show that this approach, unlike traditional methods, provides detailed insights into the strengths and weaknesses of process mining techniques, both quantitatively and qualitatively.
}

\maketitle

\section{Introduction}\label{sec:introduction}
A process mining assessment method should include \emph{validation} to check the result for correctness, quantitative and qualitative \emph{evaluation} to assess the effectiveness and accuracy, \emph{reliability} assessment to look into repeatability and reproducibility of the results, \emph{robustness} assessment to assess the ability of the method to cope with variations in the data and noise, \emph{performance} and \emph{scalability} assessment to assess efficiency and ability to cope with large and complex logs and models, and \emph{usability} assessment to evaluate interpretability of obtained results.

Already in 2008, Rozinat et al. identified the need for a common framework for evaluating process mining results~\cite{rozinat2008}. They emphasized that such a framework should help researchers compare algorithm performance and end users to validate their results. They also stressed that such a framework should allow users to influence process and log characteristics and support the generation of ``forbidden'' scenarios as a complement to the actual execution log. Despite the progress in the process mining field and the availability of tools for model generation, simulation, and injection of noise into event logs, the challenges formulated in~\cite{rozinat2008} still remain relevant.

We encountered these challenges when working on the NWO CERTIF-AI project, in which we developed alignment methods for the conformance checking of processes involving multiple entities, such as objects and resources performing different tasks, taking the interaction of these entities into account in the alignments. We obtained data from real-life processes with varying characteristics and noise levels. However, when evaluating our alignment methods to check whether the recorded behavior of a manufacturing process from our partner company Omron is compliant with a prescriptive model provided by process owners, we found it difficult to validate the results without major efforts from stakeholders and time-consuming manual inspection, making a large scale evaluation and validation practically impossible. The exposed individual deviations in process behavior can be explained in several ways by \eg timing issues in the recording, multitasking by human operators, or operators temporarily switching roles. 

To validate the results, we would need to have knowledge about the true causes of deviations, be it recording errors or behavioral outliers. To conduct a proper large-scale validation we would need multiple models with varying characteristics, multiple executions of those models with different violations of the prescribed behavior, and multiple event logs with recording errors. In addition to that we need an ``omniscient'' stakeholder, an ``oracle'' able to tell whether the deviation detected in the alignments is a true deviation and whether we provide a correct explanation of this deviation.

It is clearly infeasible to provide such a ``ground truth oracle'' when using real-life data, especially if the goal is to do experimentation on a large scale. Therefore, we resort to using synthetically generated process models and event logs with realistic characteristics. Several tools like PTAndLogGenerator~\cite{jouck2019generating}, PUPRLE~\cite{burattin2022purpose}, and AIR-BAGEL~\cite{ko2020air} make steps towards this goal. However, these inject noise directly into simulated logs, which does not allow for generating data containing typical behavioral deviations observed in real-life processes and validating whether the results of the process mining methods can handle such deviations correctly. 

In this paper, we propose an approach for generating synthetic data in which realistic noise is incorporated into a simulation model with the help of model transformations capturing patterns of both behavioral deviations and recording errors, resulting in an imperfect process model and an event log obtained for this model.
Both the transitions of the model, representing ``true'' events and deviations, and the event records in the log contain tags indicating the deviation type, thus allowing to implement an oracle aware of behavioral outliers and recording errors. 
\tent{Using this approach, we describe the creation of datasets for synthetic processes with different characteristics and illustrate the use with an alignment validation example.}

This paper is organized as follows. In Sec.~\ref{sec:related_work}, we discuss related work. In Sec.~\ref{sec:complications}, we discuss the impact of the characteristics of real-life data on the assessment and pose our research question concerning how this can be mimicked synthetically, providing knowledge of the ground truth. We describe our method and discuss how its produced components contribute to our goal in Sec.~\ref{sec:framework}, \tent{followed by a description of the current tool support in Sec.~\ref{sec:toolsupport}. Sec.~\ref{sec:datasets} demonstrates the approach describing the complete process of creating datasets of synthetic process data} and Sec.~\ref{sec:alignments} demonstrates the approach showing validation results and generated insights. We discuss the implications of our work in Sec.~\ref{sec:conclusion}.

\tent{This paper is an extension of work originally presented in ICPM~\cite{sommers2024assessing} in which we generalize the proposed approach by generating a set of event logs, instead of a single one, that can be generated from every subset of selected patterns and simulation parameters to provide a range of represented characteristics. We provide three instances of using the proposed approach resulting in three datasets of synthetic process data, each of which is characterized by recording errors and behavioral deviations that cannot be established using existing synthetic data generation methods. This illustration complemented by a description of the current status of tool support is also aimed at increasing the accessibility of our approach. Further, we elaborate on the components of the approach regarding the patterns and model transformations through examples, and we discuss the impact of the simulation parameters affecting the simulated event log.}

\section{Related Work}\label{sec:related_work}
In this section, we give a brief overview of assessment practices in process mining with real-life event logs and synthetic data. 

\subsection{Assessment using real-life logs with unknown model}\label{sec:relwork_rl}
Evaluating process mining techniques using real-life data requires significant effort and active participation of data owners. The case study with a UWV event log in~\cite{dees2017} is an excellent example of this type of work. Although such case studies provide invaluable insights and help identify strengths and limitations of process mining techniques, the proprietary nature of this data often prevents it from being shared with the process mining community for future research. The assessment of repeatability and reproducibility of the studies becomes out of reach.

An established practice for evaluating process mining techniques is to use open data sets provided in yearly Business Process Intelligence Challenges (BPIC) from 2011 to 2020, which originate from real-life processes and can be found in the \href{https://data.4tu.nl/}{4TU Data Repository}. Some of these data sets have become very popular, with BPIC 2012 being cited in nearly 500 papers and BPIC 2017 by almost 300, according to Google Scholar at the time of writing this paper. From the very first years, BPIC event data was realistically noisy in terms of both its recording and behavior~\cite{bose2013wanna}, which made this data important for testing and comparing various techniques in real-life scenarios.

A complicating factor in assessing process discovery techniques with BPIC data is that the ground truth knowledge about the underlying process is missing, with only limited information available thanks to efforts in understanding the processes through (manual) analyses and consultations with domain experts. Assessing the quality of such process mining techniques as conformance checking or model repair using BPIC data remains a complex and not straightforward task, as the corresponding behavioral models are not known. 

To address the question of whether process mining methods work on real-life small, incomplete event logs, \cite{kappel2021} developed an evaluation framework that reduces event logs and generates small event logs by removing traces either randomly or along the time dimension, producing training and test logs. Removal of traces can lead to side effects related to the workload of resources, case interactions, etc., making the log less realistic. In a sense, this approach moves towards evaluation with synthetic data.

\subsection{Assessment with synthetic data}\label{sec:relwork_synth}
Besides real-life data, synthetic data is used to provide datasets for evaluating process mining methods. A number of tools were developed for this purpose.
\emph{PTAndLogGenerator} (PTALG) described in~\cite{jouck2019generating} is a tool that allows to generates a population of non-structured process models with user-defined probabilities for sequence, choice, parallel, and loop structures. Corresponding event logs consisting of single object traces are obtained by simulation. `Noise' is imputed after simulation directly into traces and includes missing head, missing body (episode), missing tail, order perturbation, and the introduction of additional activities, adopting the definition of noise from~\cite{gunther2009process}, additionally incorporating a randomly generated decision model and data attributes.

In~\cite{burattin2022purpose}, a PURPose-Guided Log gEneration (PUPRLE) framework is proposed to produce event logs with different properties for targeting different mining purposes. It takes a process model, not restricted to its modeling language, to simulate an event log through (guided) execution of the model. Depending on the purpose of the synthetic data, the simulated event log can be attributed with different characteristics. \Eg for its use in process discovery and conformance checking, the resulting recorded behavior includes some predefined infrequent behavior, and noise, as defined by~\cite{gunther2009process}, respectively. Note that this noise is again limited to the same simple log manipulations as in PTALG, and is only imputed after simulation. Furthermore, the input process model represents the actual behavior of the process, \ie it includes behavioral outliers contributing to infrequent behavior.

AIR-BAGEL~\cite{ko2020air} is an interactive tool designed to inject pseudo-real anomalies into event logs by associating them with specific root causes, such as resource behavior or system malfunctions, using a probabilistic mechanism for resource and systems errors, resulting in e.g. skipping a step or rework. The tool generates logs augmented with labels and attributes indicating the corresponding anomaly types, enabling the evaluation of event log cleaning methods. 

%

\begin{figure}
\vspace*{-.3cm}
\centering
    \begin{subfigure}[t]{0.48\textwidth}
    \centering
    \includegraphics[scale=0.75]{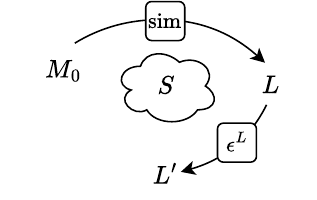}
    \caption{Synthetic process data generation methods with noise injection function $\epsilon^L$}\label{fig:synth_RW}
    \end{subfigure}
    \begin{subfigure}[t]{0.48\textwidth}
    \centering
    \includegraphics[scale=0.75]{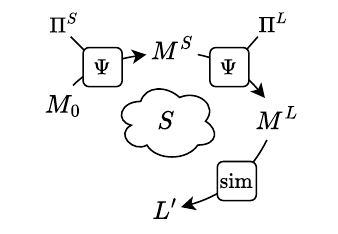}
    \caption{Generation of synthetic process data with behavioral and recording deviation patterns $\Pi^S$ and $\Pi^L$ and model transformation function $\Psi$}
    \label{fig:framework}
    \end{subfigure}
\caption{Approaches to generating synthetic data (model $M$ and log $L$) for an underlying process $S$}\label{fig:fig1}
\end{figure}

Figure~\ref{fig:synth_RW} provides a simplified illustration for these approaches where a process model $M_0$ is assumed to describe the true behavior of the process, and used to simulate a ``clean'' event log $L$. Noise is imputed only afterwards by a log manipulation function $\epsilon^L$.

The availability of such tools facilitates large-scale evaluations of process mining techniques. For example, Van Houdt et al.~\cite{vanhoudt2024} generated 400 artificial event logs using PTALG to generate models, BPSimPy to populate models with simulation parameters like the total simulation length, and L-Sim simulator to generate low-level event logs. The empirical evaluation of unsupervised log abstraction techniques presented in ~\cite{vanhoudt2024} primarily focuses on the precision and fitness of discovered models and provides interesting insights related to the influence of abstraction techniques used on the balance between fitness and precision. They note that increasing the scale of experimentation allowed them to obtain more nuanced results.

\section{Requirements to synthetic process data}\label{sec:complications}
To conduct an assessment of process mining techniques on synthetic but realistic data, we should be able to generate synthetic process data including (1) process models with different process characteristics, like the degree of parallelism or non-determinism, as they may unknowingly influence the results, (2) variation in deviations/noise, both in process execution and its recording, consistent to real-life deviation patterns, and (3) a ground truth which provides a target function for the assessment problem at hand. 

Process models representing processes with different characteristics can be generated using existing model generation tools, like PTALG. In this paper, we address requirements (2) and (3) only, assuming that we already have models representing the system, automatically or manually generated.

\begin{figure}[tb]
\centering
\hspace*{-1.8cm}
\includegraphics[scale=0.65]{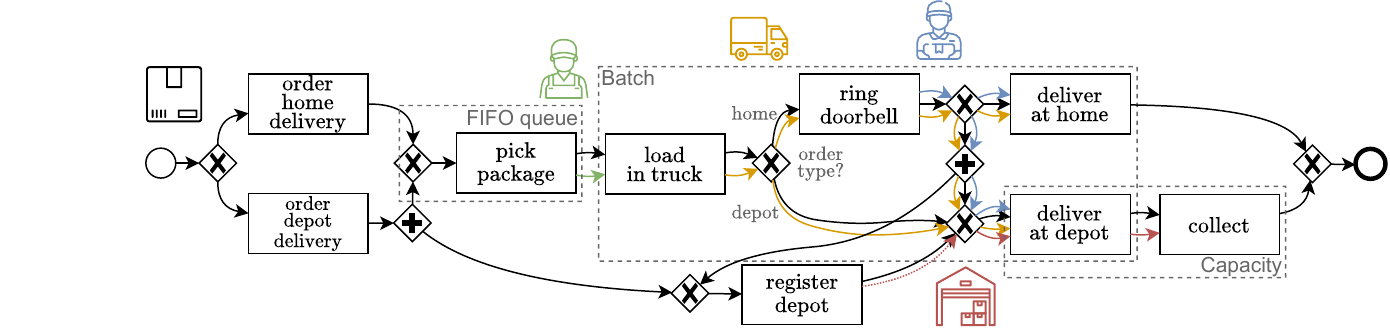}
\hspace*{1.8cm}
    \caption{\tent{Abstract process model for the package delivery process, including modeled behavior of \emph{packages} (black), \emph{FIFO queue} (yellow), \emph{warehouse employees} (green), \emph{delivery vans} (orange), \emph{deliverers} (blue), and \emph{depots} (red).}}
    \label{fig:M_package}
\end{figure}

\tent{As a running example, we manually create a synthetic package delivery process. An abstract, simplified overview of the process is illustrated in Fig.~\ref{fig:M_package}, starting with the choice of \emph{home} or \emph{depot} delivery, after which the package (black) queues for a \emph{warehouse employee} (green) to \emph{pick} and \emph{load} it into a \emph{van} (yellow). In case of home delivery, a \emph{courier} (blue) drives off and \emph{rings} a door after which he continues to either immediately hand over the package (deliver home), or deliver it at the corresponding \emph{depot} (red) after \emph{registration}, where it is left for \emph{collection}. Alternatively, for depot delivery, `ringing' and therefore also `deliver home' is omitted in the subprocess.}

\subsection{Characteristics of real-life processes and their data}\label{sec:chars}
We need mechanisms to generate data with typical \emph{recording errors} and \emph{behavioral outliers}, with the type of deviation clearly indicated for later use in the assessment.

\begin{table}[tb]
\centering
\captionsetup{width=\linewidth}
\caption{Categorization of data quality issues for each entity in recorded behavior, taken from~\cite{bose2013wanna}.
}\label{tab03:L_issues}
\begin{tabular}{|l|r|r|r|r|r|r|r|r|r|}
\hline
    & \multicolumn{1}{c|}{\rotatebox{90}{\bf{case}}} & \multicolumn{1}{c|}{\rotatebox{90}{\bf{event}}} & \multicolumn{1}{c|}{\rotatebox{90}{\bf{belongs to}}} & \multicolumn{1}{c|}{\rotatebox{90}{\bf{c\_attribute}}} & \multicolumn{1}{c|}{\rotatebox{90}{\bf{position}}} & \multicolumn{1}{c|}{\rotatebox{90}{\bf{activity name~}}} & \multicolumn{1}{c|}{\rotatebox{90}{\bf{timestamp}}} & \multicolumn{1}{c|}{\rotatebox{90}{\bf{resource}}} & \multicolumn{1}{c|}{\rotatebox{90}{\bf{e\_attribute}}} \\ \hline

Missing data &

\textcolor{black}{\Lissue{mi}{c}} & \Lissue{mi}{e} & \Lissue{mi}{o}   & \textcolor{black}{\Lissue{mi}{ca}} & \Lissue{mi}{p}   & \textcolor{black}{\Lissue{mi}{a}} & \textcolor{black}{\Lissue{mi}{t}} & \Lissue{mi}{o}   & \textcolor{black}{\Lissue{mi}{ea}}
\\ \hline

Incorrect data &

\textcolor{black}{\Lissue{in}{c}} & \Lissue{in}{e}  & \Lissue{in}{o}   & \textcolor{black}{\Lissue{in}{ca}}  & \Lissue{in}{p}   & \Lissue{in}{a} & \textcolor{black}{\Lissue{in}{t}} & \Lissue{in}{o}   & \textcolor{black}{\Lissue{in}{ea}} 
\\ \hline

Imprecise data  &

& &  \textcolor{black}{\Lissue{im}{o}}  & \textcolor{black}{\Lissue{im}{ca}}  & \textcolor{black}{\Lissue{im}{p}}   & \textcolor{black}{\Lissue{im}{a}} & \textcolor{black}{\Lissue{im}{t}} & \textcolor{black}{\Lissue{im}{o}}   & \textcolor{black}{\Lissue{im}{ea}} 
\\ \hline
Irrelevant data & 

\textcolor{black}{\Lissue{ir}{c}} & \textcolor{black}{\Lissue{ir}{e}}  &&&&&&&  \\ \hline








\end{tabular}
\end{table}

We consider recording errors in event logs as records misrepresenting the actually executed behavior of the process. In~\cite{bose2013wanna}, \emph{recording errors} are subdivided into four categories of issues, where data could be (1) missing, (2) incorrect, (3) imprecise, and (4) irrelevant. Each category could apply to various elements of the event log, \eg events, event attributes, relations between attributes and events, and between events themselves. Tab.~\ref{tab03:L_issues} summarizes these quality issues. Possible sources of recording errors are faulty logging mechanisms, either automatic or manual, with unsynchronized clocks, too coarse timestamps, data corruption, and filtering and aggregation methods.

\tent{Recent advancements~\cite{basmer2024classification} extend these quality issues to object-centric event data, further classifying the issues on object-centric properties like object types and their relations. While this fits the processes we consider in this work, the essence of the data quality issues is the same. All issues from Tab.~\ref{tab03:L_issues} are represented in the set of processes from the BPIC datasets~\cite{bose2013wanna}, and could be reflected in the running example of the synthetic package delivery process. For the examples that follow we focus primarily on the categories of \emph{missing} and \emph{incorrect} data.}

A process model ties together a set of modeling patterns~\cite{russell2006workflow} to define the ``main'' behavior of the process. The interpretation of the ``main behavior'' depends on the process mining task. It could be the expected behavior, the ``happy flow'' behavior, or frequent behavior. \emph{Behavioral outliers} refer to behavior that deviates from the \emph{main behavior} of the process. There could be several reasons for violating the modeled behavior, \eg people not following guidelines, or encountering rarely seen cases not taken into account in the model, thus choosing different behavioral patterns than the ones imposed by the model.

\tent{The abstract process model from Fig.~\ref{fig:M_package} includes various modeling patterns, like choices, parallelism, queueing, batching, capacity, resource memory, and (continuous) correlation, involving several objects simultaneously. Behavioral deviations are considered patterns as well, which in essence boil down to similar issues as the recording errors when they are captured in the event log. Later, we explicitly draw connections between the two types of issues to signify this relation.}

Process mining techniques should be robust to recording errors and should be able to handle behavioral deviation appropriately (defined \eg based on their frequencies), depending on the use case. Take for example process discovery. When the discovered process model is to be used as a handbook for employees, the goal might be to capture only the frequent behavior, whereas when the process model is to be used as a digital twin, the goal is to capture all possible behavior.

When resorting to the use of synthetic data, these deviation characteristics should be mimicked as closely as possible to represent real-life data. Model-log pairs that can be obtained by the approach illustrated in Fig.~\ref{fig:synth_RW} are not sufficient, as (1) the model of the ``real system'' is not provided: the base model does not show behavioral deviations, or they are included in the base model and not distinguishable from regular behavior, (2) the log manipulations are performed irrespectively of the behavior of the process, therefore losing the link between model and log.

\subsection{The need for ground truth}
Dealing with these categories of errors and deviations is a challenge for process mining methods in real-life settings in general. This holds even more so for their assessment, as the explanations for causes of deviations may be ambiguous or missing altogether. For the evaluation of process discovery techniques, conformance metrics allow for measuring the quality of a process model where the ground truth model is unknown. However, this assumes the event log is free of recording errors, since differentiating them from behavioral outliers may be impossible due to ambiguity in their causes. While recording errors should ideally be filtered out completely, (some) behavioral outliers should end up in the behavior of the discovered process model. Therefore, for validation, being able to make this distinction is required. Similarly, for conformance checking, assessing whether deviations are correctly exposed and explained, having knowledge of the causes is essential for constructing the target. 

Without going into detail, we argue that other PM problems, \eg decision mining, model or log repair, bottleneck detection, and predictive and prescriptive process monitoring, suffer from the same complications.
Therefore researchers are forced to resort to assessing PM techniques using synthetically generated data.

This brings us to our research question (RQ):\\
\\
\emph{RQ: How can we set up an experiment with realistically noisy process data where the ground truth is known to allow for a complete assessment?}\\

\section{Generating Process Data with Ground Truth}\label{sec:framework}
Depending on the goal of a PM algorithm/tool, and its input and output, its assessment may require a process model, an event log, and the ground truth containing information about both behavioral and recording deviations that are to be used in assessing the quality of the produced results.

In this section, we propose an evaluation framework for generating such process data using a collection of deviation patterns and corresponding model transformations to build in the deviation into an initial process model. We first give a high-level overview of the framework, after which we discuss deviation patterns and model transformations.

\subsection{Framework design}\label{sec:framework_design}

Our framework is schematically depicted in Figure~\ref{fig:framework}.
$M_0$ denotes a base model representing the expected behavior of the process, $M^S$ serves as the true model of the process, capturing behavioral deviations, and $M^L$ extends this model by imitating faulty logging mechanisms in the model. A simulation of $M_0$  results in an ``exemplary'' log showing the expected behavior of a process. A simulation of $M^S$ gives a ``clean'' event log, showing the ``real'' behavior of a process. Simulating $M^L$ results in log $L'$ capturing the real behavior with realistic recording errors. This log $L'$ is the target of our framework.

$M_0$, $M^S$, $M^L$, $L'$ together with the links between them provide the ground truth about the process and its execution, including behavioral characteristics of the process and recording characteristics of the event log as well as its completeness with respect to the process ($M^S$) and the model ($M_0$).

The base model $M_0$ might describe an existing process, be a hypothetical process designed manually with certain characteristics in mind, or it can be generated automatically by a model generation tool. Recall that $M_0$ merely serves as a base process that does not represent the ``real'' process execution. It can serve as \eg a normative model in the assessment of conformance checking techniques, or a ``typical flow'' model in the assessment of process discovery techniques.

We start with a base process model $M_0$ and a set $\Pi$ of deviation patterns. Deviations, either recording or behavioral, are modeled as templates. A model transformation function $\Psi$ defines how the behavior modeled in deviation pattern $\pi \in \Pi$  can be added to a model. The deviating behavior is incorporated into the base process model, resulting first in model  $M_S$ of process executions with behavioral deviations, and subsequently in model $M_L$ incorporating log recording errors on top of that. By simulating the resulting model $M_L$, we acquire a realistic event log $L'$ and the knowledge of all deviations that took place, since the simulated events can be linked back to transition firings from $M^L$, which in their turn can be linked to the transitions of $M_S$ and $M_0$, or are labeled as corresponding deviations.

The set $\Pi$ of deviations consists of subsets $\Pi^S$ with behavioral deviation patterns and $\Pi^L$ with recording error patterns. A model transformation function $\Psi$ takes a model $M$, a deviation pattern $\pi\in \Pi$, and a function $h$ mapping elements of $\pi$ to the elements of $M$ in order to indicate where the deviation should occur in the model. The model transformation series starts with $M_0$, and after applying the first transformation, applies the next transformation to the obtained $\Psi(M_0, \pi_1, h_1)$, etc., finally leading to the model $M^S$ of the actual process execution. Similarly, we start with $M^S$ to apply patterns from $\Pi^L$ and obtain $M^L$ after a series of model transformations. When simulated, $M^L$ produces the event log $L'$, encoding the complete information about the transitions that generated events.

Only partial information on the events (\eg transition labels) is given to a process mining technique to be assessed. The rest of the information constitutes the ground truth knowledge to be used to compute quality metrics in the assessment and obtain qualitative insights.

\subsection{Usage in assessments}\label{sec:usage}
To use the evaluation framework, assessors must define a ground truth function $\gt$ for their PM problem and assessment tasks. This function transforms the combination of  $M_0$, $M^S$, $M^L$, and $L'$ to the target result for that process mining technique. Further on, the assessors are to choose a distance function $d$ to compare the result produced by the process mining technique to the ground truth target. Generically,
\begin{equation*}
    d\left(f(M_0,L'), \gt^f(M_0,L',M^S,M^L)\right)
\end{equation*}
measures the quality of a PM method $f$, using distance function $d$ and ground truth function $\gt^f$.

Evaluation and validation of \emph{process discovery} (PD) and \emph{process repair} techniques can be performed by comparing discovered models to either $M_0$, $M^S$, and/or $M^L$, depending on the application at hand. 
For example, the goal can be to discover an interpolation of $M_0$ and $M^S$, including frequent behavioral deviations but ignoring the infrequent ones and all the recording errors. The target process model is defined by $M_0$ and a subset  ${\Pi^S}' \subseteq \Pi^S$ of behavioral deviations. The similarity between a discovered model $PD(L')$ and $\gt^{PD}(M_0,M^S,M^L,L')$ can be computed \eg using model similarity~\cite{dijkman2011similarity}.

The goal of \emph{conformance checking} (CC) with $f=CC$ is to identify and explain deviations in the recorded behavior from the behavior captured by a normative model. A CC technique takes $M_0$ and $L'$ as input e.g. to compute an alignment of the actual behavior of the process and the modeled behavior. The ground truth target can be defined as the optimal alignment, interpreting events corresponding to the firing of original transitions of $M_0$ as synchronous moves and events corresponding to the transitions being part of the deviation patterns as log moves or model moves accordingly.

The similarity between the alignment $CC(M_0, L')$ produced by a conformance checking technique and the ground truth alignment $\gt^{CC}(M_0, M^S, M^L, L')$ can be computed \eg by sequence or graph edit distance, depending on the formalization used to model alignments~\cite{yujian2007normalized,gao2010survey}. Note that this is an abstract definition of $\gt^{CC}$. We concretize this in Sec.~\ref{sec:alignments}, where we use our proposed evaluation framework to evaluate the explainability of multi-object alignments~\cite{sommers2022aligning} and relaxed multi-object alignments~\cite{sommers2024conformance}.

\subsection{Deviation patterns}\label{sec:deviation_patterns}
\begin{figure}[tb]
    \centering
    \captionsetup{width=\linewidth}
    \captionof{table}{Deviation patterns.}\label{tab:devpatterns}
    \includegraphics[scale=0.8]{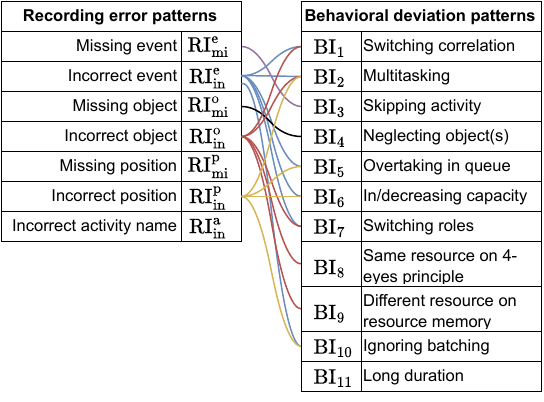}
    \vspace*{-0.5cm}
\end{figure}

In Tab.~\ref{tab:devpatterns}, we list a selection of deviation patterns that are currently implemented in our framework. The left-hand columns form the set of issues from Tab.~\ref{tab03:L_issues} specified for recording errors, originating from the issues presented in~\cite{bose2013wanna}. The right-hand columns present a concrete set of behavioral issues, originating from violations of the modeling patterns discussed in~\cite{russell2006workflow}.
As discussed in Sec.~\ref{sec:chars}, these issues can be similarly categorized as the recording issues. The counterparts are denoted by the colored links between the columns. \Eg for behavioral deviations, there can be several deviations that fall under `incorrect object' ($RI^o_{in}$), namely switching correlation, multitasking, etc.

Take for example the recording error pattern $RI_{mi}^e$, describing a missing event, where an activity is executed but failed to be recorded. Skipping an activity during the execution of a process ($BI_1$) is a behavioral deviation being a counterpart of a missing event: In both cases, the corresponding event will be missing in log $L'$, although the causes of that would differ.

A deviation pattern consists of a behavioral description and an abstract process model fragment. The abstract process model fragment serves as a blueprint for incorporating the deviating behavior into a process model.
Formally, a \emph{deviation pattern} is an abstract process model $\bar{M}$ describing deviating behavior in relation to some process behavior. {Abstract} means here that certain model elements are represented by wildcards, denoted as $\wc{*}$, acting as placeholders for specific elements of the model to which the deviation is to be added.

\begin{figure}[tb]
\centering
\hspace*{-1.8cm}
\includegraphics[scale=0.7]{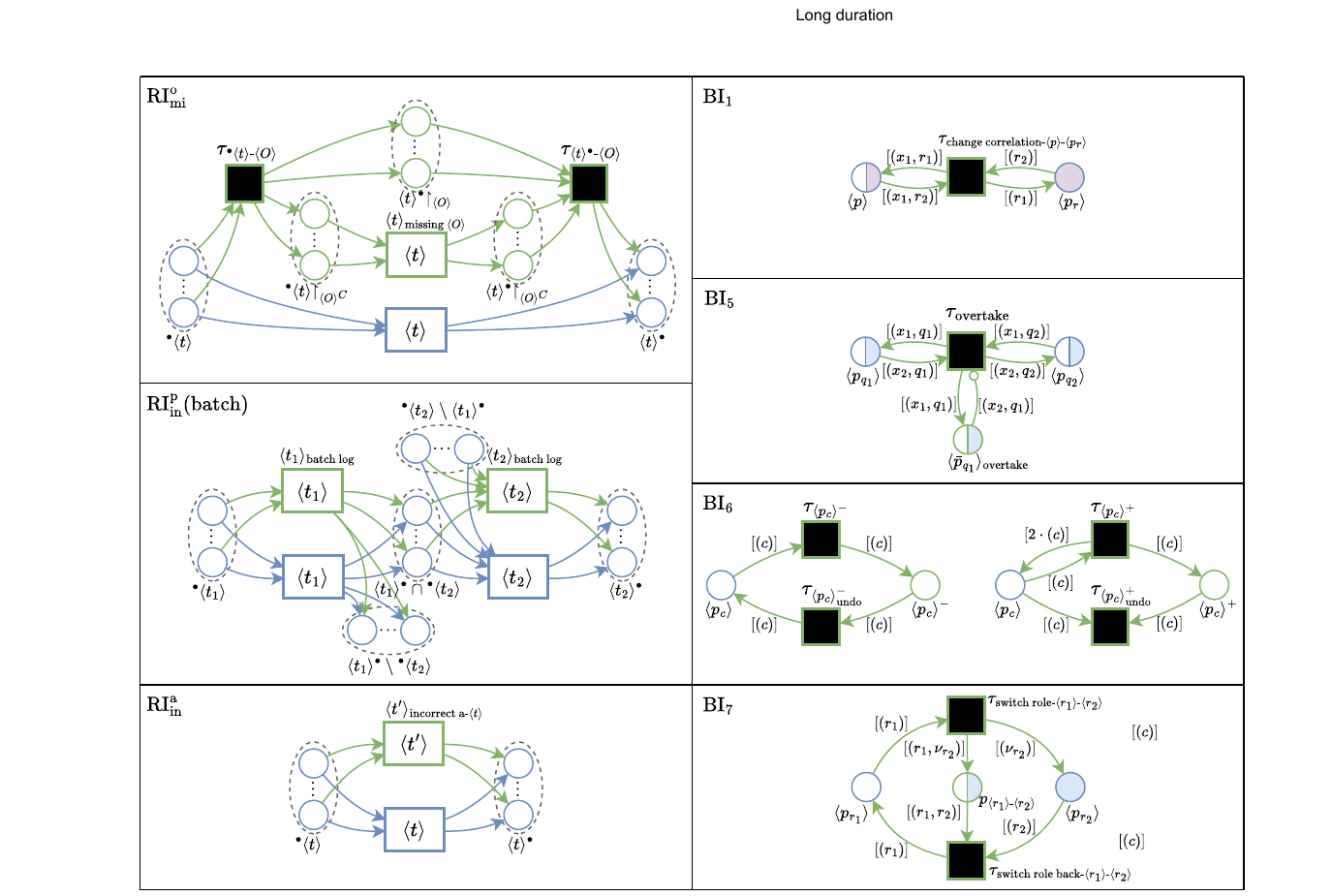}
\hspace*{1.8cm}
    \caption{\tent{t-PNID modeling blueprints for three recording issue patterns (left) and four behavioral deviation patterns (right). Blue and green elements respectively correspond to matched and newly created elements.}}
    \label{fig:patterns}
\end{figure}
To illustrate and clarify the use of patterns, we elaborate on the modeled deviation patterns for a few examples from Tab.~\ref{tab:devpatterns}, using typed Petri nets with Identifiers (t-PNIDs)~\cite{van2022data} as the modeling formalism. This formalism allows for modeling intricate behavior of several interacting objects, corresponding to patterns described in Sec.~\ref{sec:complications}. \tent{A t-PNID consists of transitions (rectangles), typed places (colored circles), and labeled arcs. A transition can consume and produce tokens by `being fired', respectively from its incoming places, \ie with an incoming arc from a place, and to its outgoing places, which are denoted as the transition's \emph{pre-set} and \emph{postset}. Tokens in a t-PNID carry (multiple) identifiers representing the object names in the system. The firing of a transition represents an event with the involved objects corresponding to the objects represented by the consumed and produced tokens and potentially enables new transition firings which determines the positions in their ordering. For the formal semantics of t-PNIDs, we refer to~\cite{van2022data}.}

\tent{Fig.~\ref{fig:patterns} shows examples of abstract modeling blueprints for the following patterns:
\begin{itemize}
    \item[$RI_{mi}^o$] The recording error of a \emph{missing object(s)} can be modeled by the abstract t-PNID shown on the top-left of Fig.~\ref{fig:patterns}. Transition $\wc{t}$ and its pre- and post-set with corresponding arcs (blue) are matched from the base model. Assuming that the place types coincide with the types of the object(s) $\wc{O}$ to be missing, a second transition $\wc{t}_{\text{missing }\wc{O}}$ is created (green) with the same activity name as $\wc{t}$ but without the involvement of $\wc{O}$. To achieve this behavior, two additional silent transitions $\tau_{\pre{\wc{t}}\text{-}\wc{O}}$ and $\tau_{\post{\wc{t}}\text{-}\wc{O}}$ are created, which bypass the tokens from places corresponding to objects from $\wc{O}$, \ie from $\pre{\wc{t}}$ to $\post{\wc{t}}\proj_{\wc{O}}$, and subsequently to $\post{\wc{t}}$. Tokens from the other places, \ie $\pre{\wc{t}}\proj_{\wc{O}^C}$ and $\post{\wc{t}}\proj_{\wc{O}^C}$, are passed on to $\wc{t}_{\text{missing }\wc{O}}$ where $\wc{O}^C$ denotes the complement set of $\wc{O}$.
    \item[$RI_{in}^p$] The recording error of \emph{incorrect position} occurring due to batch logging can be modeled by the abstract t-PNID shown on the middle-left of Fig.~\ref{fig:patterns}. This pattern is designed to be applied to a specific batching pattern in the base model (blue), where a first transition ($\wc{t_1}$) initializes work on a batch of objects, and a second transition ($\wc{t_2}$) processes the objects one-by-one. The created elements, \ie $\wc{t_1}_\text{batch log}$ and $\wc{t_2}_\text{batch log}$ with the same activity names as $\wc{t_1}$ and $\wc{t_2}$ represent the same behavior. However, by applying different timing specifications, the simulation shows the behavior where $\wc{t_1}_\text{batch log}$ takes as long as processing the whole batch and $\wc{t_1}_\text{batch log}$ takes a negligible amount of time, representing the same activities but logged at an incorrect position. The distinction between the intersection and differences between the post-set of $\wc{t_1}$ and the pre-set of $\wc{t_2}$ is made to correctly pass along the tokens.
    \item[$RI_{in}^a$] The recording error of an \emph{incorrect activity name} corresponding to an event can be modeled by the abstract t-PNID shown on the bottom-left of Fig.~\ref{fig:patterns}. This is a simple pattern, creating a duplicate transition $\wc{t'}_{\text{incorrect a-}\wc{t}}$ for a matched transition $\wc{t}$ inheriting its pre- and post-set and with a different activity name, specified by $\wc{t'}$.
    \item[$BI_1$] The behavioral deviation of \emph{changing correlation} can be modeled by the abstract t-PNID shown on the top-right of Fig.~\ref{fig:patterns}. It consists of two matched places, $\wc{p}$ and $\wc{p_r}$ (blue) and one created silent transition $\tau_{\text{change correlation-}\wc{p}\text{-}\wc{p_r}}$ (green), with wildcards for $\wc{p}$ and $\wc{p_r}$. $\wc{p}$ is assumed to be a place in the base model with two types, of which one corresponds to a resource with the same type as $\wc{p_r}$. The silent transition, through the labels on the connected arcs, consumes a correlation token from $\wc{p}$ and a resource token from $\wc{p_r}$ to change the correlation to the new resource. In~\cite{sommers2024assessing}, a similar pattern, falling under the same behavioral deviation, is presented which swaps the resources from two existing correlation tokens instead of changing the correlation for one token.
    \item[$BI_5$] The behavioral deviation of \emph{overtaking} in a first-in-first-out queue can be modeled by the abstract t-PNID shown on the upper-middle-right of Fig.~\ref{fig:patterns}. A created silent transition $\tau_\text{overtake}$ swaps the correlation from two objects in matched places $\wc{p_{q_1}}$ and $\wc{p_{q_w}}$ with their respective queue objects denoting the place in the queue, similarly as the previous pattern. Additionally, a created place $\wc{p_{q_1}}_\text{overtake}$ is added to prevent a continuous cycle of objects overtaking each other in the simulation of the model.
    \item[$BI_6$] The behavioral deviations of \emph{decreasing} and \emph{increasing} the capacity of a resource object can be modeled by the abstract t-PNID shown on the lower-middle-right of Fig.~\ref{fig:patterns}. A created silent transitions $\tau_{\wc{p_c}^-}$ decreases the capacity of an object in matched placed $\wc{p_c}$, stores this information in a created place $\wc{p_c}^-$ which can subsequently be undone by silent transition $\tau_{\wc{p_c}^-_\text{undo}}$. Similarly, to increase the capacity, $\tau_{\wc{p_c}^+}$ duplicates a token in $\wc{p_c}$, which can be undone by $\tau_{\wc{p_c}^+_\text{undo}}$ which consumes the duplicated tokens.
    \item[$BI_7$] The behavioral deviation of \emph{switching roles} of a resource object can be modeled by the abstract t-PNID shown on the bottom-right of Fig.~\ref{fig:patterns}. Two matched places $\wc{p_{r_1}}$ and $\wc{p_{r_2}}$ correspond to places where resource objects with two different roles reside. A created silent transition $\tau_{\text{switch role-}\wc{r_1}\text{-}\wc{r_2}}$ can move a token representing a resource object from role $\wc{r_1}$ to the role of $\wc{r_2}$. Similarly to the previous pattern, this information is stored in place $p_{\wc{r_1}\text{-}\wc{r_2}}$ to undo this move with created silent transition $\tau_{\text{switch role back}\wc{r_1}\text{-}\wc{r_2}}$. Note that the $\nu$ variables in the arc labels ensure that the objects from different types are non-overlapping, by synthesizing a new identifier that references the corresponding object from role $\wc{r_1}$.
\end{itemize}}

\tent{We refer to~\cite{sommers2024assessing} for four more examples of the patterns $RI_{mi}^e$, $RI_{in}^e$, $BI_3$, $BI_2$, and $BI_1$ in which two correlations are swapped, \ie slightly different from $BI_1$ as presented here.}

Our approach is defined conceptually, independent of the modeling formalism, as long as the formalism of the abstract deviation patterns, both for behavioral outliers and recording errors, matches that of the base model. Therefore, it is not restricted to the use of t-PNIDs.

The list can be extended with additional pattern descriptions and corresponding definitions. All provided patterns are modeled in a way that the model transformation is additive in behavior, \ie behavior described in the pattern is added to the base model without eliminating any behavior that was captured in the model. For this reason, multiple deviation patterns do not interfere with each other. The created elements have distinct names. This is a necessary property to ensure that the resulting process model allows sufficient freedom to define ground truth functions for different process mining tasks.

\subsection{Model transformations}\label{sec:model_transformations}
Behavior from a deviation pattern $\pi$, described by an abstract process model $M_\pi$, can be incorporated into a process model $M$, through a model transformation function $\Psi$ which uses $M$, $M_\pi$, and an injective mapping function $h$ mapping the wildcards from $\pi$ to elements in $M$.

Our definition for $\Psi$ is derived from Petri net transformations as discussed in~\cite{ehrig2003graph} and is defined as follows:
\begin{equation}
    \Psi(M,\pi,h) = M \cup h(\pi)
\end{equation}
where $h(\pi)$ is a morphism from the abstract process model fragment $\pi$ to a concrete process model fragment in the space of $M$, with \emph{matched elements} referring to elements in $M$. Through $\Psi$, $\Psi(M,\pi,h)$ additionally contains the \emph{created components} of $\pi$, providing the added behavior.

\begin{figure}[tb]
\centering
\hspace*{-1.1cm}
\includegraphics[scale=0.65]{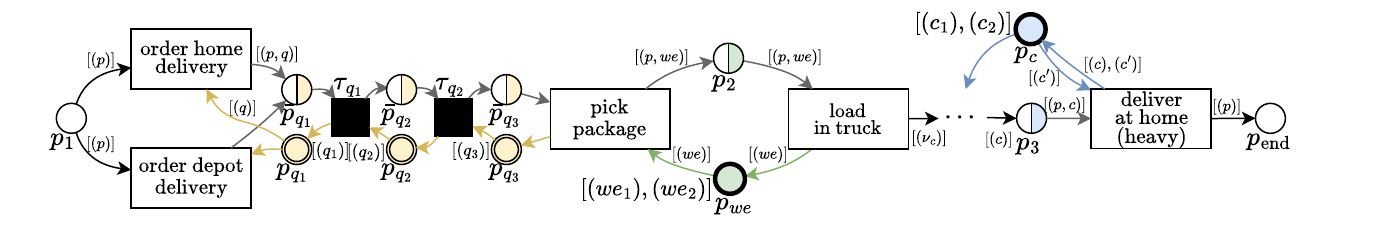}
\hspace*{1.1cm}
    \caption{\tent{t-PNID fragment $M'$ of a modified version of $M$, with the addition of a heavy delivery at home.}}
    \label{fig:M_package_}
\end{figure}

\begin{figure}[tb]
\centering
\hspace*{-1.2cm}
\includegraphics[scale=0.65]{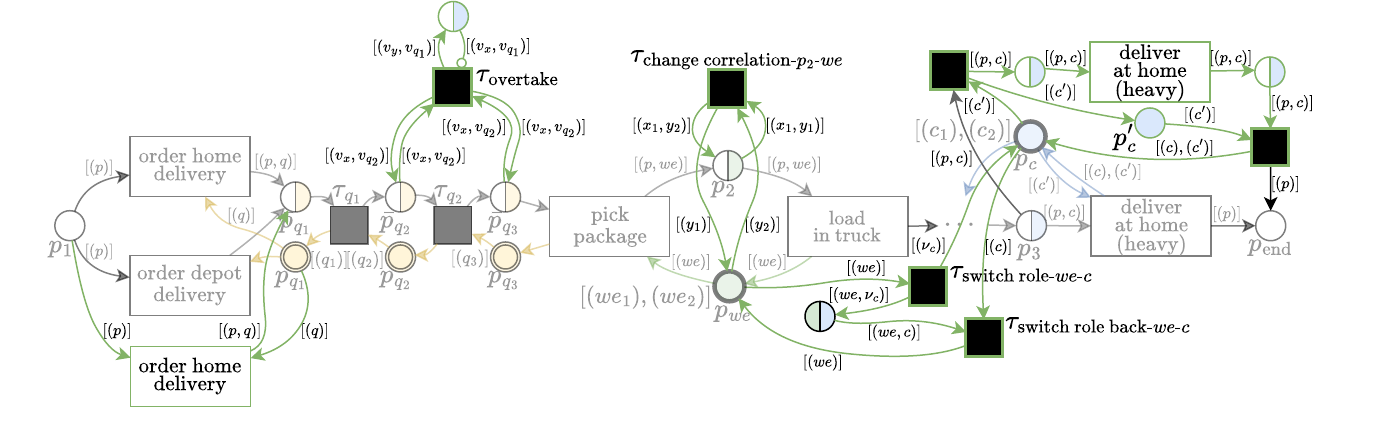}
\hspace*{1.2cm}
    \caption{\tent{The result of model transformations applied to $M'$.}}
    \label{fig:mt}
\end{figure}

\tent{To illustrate how patterns from the examples in Fig.~\ref{fig:patterns} are applied to a process model, we show in Fig.~\ref{fig:M_package_} a fragment of the t-PNID implementation $M'$ of a slightly modified version of the abstract process model from Fig.~\ref{fig:M_package} for the running example. This is the same formalism as the examples presented for some of the recording error and behavioral deviation patterns. Additionally, not shown in the patterns, the process' initial state is denoted by the multisets of object names ($\mset{(we_1),(we_2)}$ and $\mset{(c_1),(c_2)}$) in places with a thicker outline, \ie $p_{we}$ and $p_c$, modeling warehouse employees and couriers. To apply the pattern for a missing recorded object, the process is extended with an additional activity for \emph{delivery of heavy packages at home}, requiring two couriers for its execution.}

\tent{Fig.~\ref{fig:mt} depicts the applied model transformations of four patterns, \ie \emph{overtaking} ($BI_5$), \emph{changing correlation} ($BI_1$), \emph{switching roles} $BI_7$, and \emph{missing recorded objects} ($RI_{mi}^o$), to $M'$. The pattern $BI_5$ for overtaking from Fig.~\ref{fig:patterns} is applied with mapping $\set{\wc{p_{q_1}} \mapsto \barp_{q_2}, \wc{p_{q_2}} \mapsto \barp_{q_3}}$, to create a silent transition connecting to these existing places from the base model $M'$ which can be used to swap the correlation between two objects and their place in the queue. Similarly, the pattern $BI_2$ for changing correlation from Fig.~\ref{fig:patterns} is applied to base model $M'$ with mapping $\set{\wc{p} \mapsto p_2, \wc{p_r} \mapsto p_{we}}$ where $p_{we}$ is the resource place for warehouse employees. The created silent transition can be used to change the correlation of the employee loading a package in a truck after it is picked by a different employee. The pattern $BI_6$ for switching roles is applied $M'$ to temporarily switch the role with the created silent transition of a warehouse employee to that of a courier, by the mapping to the corresponding resource places, \ie $\set{\wc{p_{r_1}} \mapsto p_{we}, \wc{p_{r_2}} \mapsto p_c}$.}

\tent{Lastly, the pattern $RI_{mi}^o$ is applied $M'$ to model the recording of a heavy delivery with a missing second courier. Note that through the created silent transition, both couriers are claimed throughout the execution of the created heavy delivery transition, which however is unaware of the second involved courier, as it is bypassed through the created place $p_c'$.}

\tent{Note that there are restrictions on the mapping of the wildcards to correctly model the behavioral pattern into the base model. Places and transitions have specific roles in the net, which should match the roles of the to-be-matched elements from a pattern. This could be either checked in the model transformation, or the user should be instructed clearly on the requirements of the wildcards. \Eg the multitasking pattern is defined behaviorally that the correlation is prematurely destroyed, putting a resource from busy to idle. This requires that $\wc{p_1}$ and $\wc{p_2}$ are mapped respectively to a correlation/busy place of a resource role and its corresponding idle place.}

\subsection{Simulation}\label{sec:simulation}
The event log $L'$, which is the last element generated by the framework, describes the recorded behavior of the process. $L'$ is obtained by discrete event simulation of the process model $M^L$, which represents the real behavior of a process including behavioral outliers (by $\Pi^S$) and faulty recording mechanisms (by $\Pi^L$), deviating from the expected behavior modeled in $M_0$.

The simulation module is an interchangeable part of the framework. \Eg $L'$ can be obtained by a play out of $M^L$, allowing for repeated firings of transitions, either through being enabled, or scheduled in the form of arrivals or predetermined schedules. 

\tent{For a more advanced simulation method, information about stochastics and timing could be included in the process model. Example modeling formalisms are Stochastic Petri nets~\cite{leemans2021stochastic} and Timed Petri net~\cite{razouk1983performance,zuberek1980timed}. We explain the concepts here without going into the formal definition of these formalisms.}

\tent{Adding weights to transitions of a net provides probabilities to transition firings. A transition firing can be sampled from a categorical distribution for all enabled transitions with corresponding bindings/modes, with probabilities defined by the transition weights. For example, with enabled transition firings $T_\mode^e$ and weight function $w: T \rightarrow \Real^+_0$, the probability that the selected transition firing, denoted as $t_\mode^s$, is $t_\mode \in T_\mode^e$ is defined by
\begin{equation}
    p(t_\mode^s = t_\mode | w) = \frac{w(t)}{\sum_{t_\mode' \in T_\mode^e} w(t')}
\end{equation}}

\tent{A transition firing consumes tokens from incoming places and produces tokens in outgoing places. Until now, this has been defined atomically. Delaying the production of tokens adds time specifications and durations to the simulation of the execution of the process model. The duration of the delay is specified on transitions or, more granularly, on outgoing arcs of transitions.
Transitions are withheld from consuming tokens during the delays of the corresponding production in places. There are various ways to define the time specification, \eg by constants, functions, or distributions, as long as the (sampled) delay is non-negative. Let us say for example that the time it take from picking a package and loading it in a truck is normally distributed with a mean and variance of respectively of fifteen and six minutes. The duration can be specified by $+\mathcal{N}(15, 6)$ on the arc from transition `pick package' to place $p_2$ in the t-PNID model depicted in Fig.~\ref{fig:M_package_}.}

\tent{Both the transition weights and time specifications allow for additional conditional parameters, by adding a variable denoting \eg the current marking of the net, the \emph{current simulation time} $\eta \in \Time$, or other data attributes~\cite{holliday1987generalized}. With $w: T \times \Time \rightarrow \Real^+_0$, specific transitions can be temporarily disabled, which is especially useful for controlling probabilities and frequencies of deviating behavior during simulation. Similarly, for time specifications, \eg the variance of the resources' productivity can be more accurately modeled based on the current simulation time.}

\tent{In Sec.~\ref{sec:deviation_patterns}, we define deviation patterns, both for behavioral outliers and recording errors, on the behavior of the modeling formalism. Note that with extended modeling formalisms for more advanced simulation, like SPNs and TPNs, deviations can be naturally extended to the stochastics of the net. Two examples of this are the pattern $RI_{in}^p$ for batch logging as described in Sec.~\ref{sec:deviation_patterns} and $BI_{11}$ for long durations from Tab.~\ref{tab:devpatterns}. In both patterns, the behavior does not change from the perspective of the workflow, however, the timing specifications characterize the deviating behavior during simulation. As such, deviations on the stochastics, timings, and other external data attributes can be modeled directly in the sampling distributions for weights and durations based on conditionals like time or markings.}

Independently of the specific simulation module, the simulated behavior of the process model $M^L$ resembles real-life behavior by design, as it incorporates both behavioral deviations and recording errors.

\section{\tent{Tool support}}\label{sec:toolsupport}
\tent{We provide two versions of implementations of the evaluation framework, as proposed conceptually above. One version is a tool, called Trident, with a graphical user interface~\cite{sommers2023trident}, offering a user-friendly interaction but is limited in functionality. This version can be operated through either a Flask GUI, a command line interface, and/or run from other Python code, to generate ground truth synthetic process data. The source code, an installation manual, and a screencast of example usage are available at \href{https://gitlab.com/vignesh_dv/mira/-/tree/paper/mira/pattern}{gitlab.com/vignesh\_dv/mira/-/tree/paper/mira/pattern}.}

\tent{The other version contains the source code implementation with example usage for more advanced deviation patterns and simulation with handling of stochastics and time durations~\cite{sommers2024assessing}. Also based on Python, this version can be operated by creating scripts invoking the appropriate classes and definitions. The source code and usage instructions are available at \href{https://gitlab.com/dominiquesommers/mira/-/tree/main/mira/simulation}{gitlab.com/dominiquesommers/mira/-/tree/main/mira/simulation}.}

\tent{Both implementations operate on regular Petri nets~\cite{peterson1981petri,murata1989petri}, \nunets~\cite{rosa2010decision}, Resource-Constrained \nunets~\cite{sommers2022aligning,sommers2023exact} (RC \nunets), and t-PNIDs, and can be extended to other formalisms. One of which is a generalization of t-PNIDs for handling variably involved objects for modeling one-to-many interactions, similarly as Object-centric Petri nets with Identifiers (OPIDs)~\cite{gianola2024object}, Object-centric Petri nets (OC nets)~\cite{van2020discovering}, and Synchronizing proclets~\cite{fahland2019describing}. The framework can therefore also be used to generate logs of object-centric processes.
This allows to generate rich logs of processes with event records including information about resources that participated in the activities' executions.}

\tent{The user interface of Trident is focused towards resource-constrained \nunets~\cite{sommers2022aligning}, distinguishing between the place types being regular, resource available, and resource busy. Recall from Sec.~\ref{sec:model_transformations} that the model transformations put requirements on proper mapping of elements from the abstract deviation pattern to the base model. These requirements can be easily checked, due to the strictly defined modeling restrictions of RC \nunets.}

\tent{In general, the usage for both tools is similar and follows the concept as described in Sec.~\ref{sec:framework_design}. \Ie one takes a real-life process model or sketches a hypothetical process and a set of appropriate deviations. The deviation patterns from Tab.~\ref{tab:devpatterns}, are provided and already defined. Not being restricted to those, additional patterns can be defined to see fit. The selected deviations are applied to the base model through a series of model transformations, and the resulting process model $M^L$ can be either simulated directly or alternatively exported for usage in an external simulation module.}

\tent{The simulation module included in Trident is a simple play out of $M^L$ from the provided initial to final marking, with the option to set a limit of the number of transition firings, in case of infinite behavior. Probabilities are modeled through sampling a waiting time for transition firings from the moment they are enabled. In case the transition is not enabled anymore at the scheduled time, its firing is canceled. The simulation is basic in terms of probabilities of which transitions to fire, \ie it does not take into account other dependencies than the sampled scheduling time from the moment it is enabled.}

\tent{The more advanced implementation provides a simulation module with sampling of transition firings and duration of token productions as described in Sec.~\ref{sec:simulation}. Furthermore, the arrivals of spontaneous objects and the start/stop schedules of scheduled objects can be defined by sampling distributions. The conditional variable of the current simulation time can be used to \eg vary the probability of deviating behavior over time.}

\tent{If one requires a more advanced simulation, an export of process model $M$ can be used in other tools.}

\section{\tent{Putting the framework into practice}}\label{sec:datasets}

\begin{figure}[tb]
\centering
\includegraphics[scale=0.65]{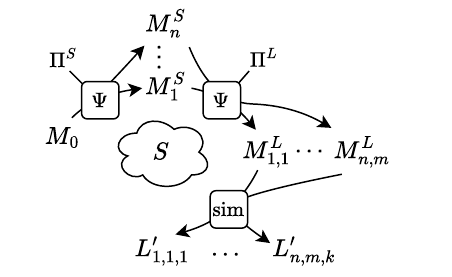}
    \caption{\tent{Framework extension to generate multiple event logs from a single base model.}}
    \label{fig:framework_extension}
\end{figure}

\tent{Starting from a single base model $M_0$ for a (synthetic) process, the framework, as described in Sec.~\ref{sec:framework_design} and illustrated in Fig.~\ref{fig:framework}, can be used to generate a versatile dataset of event logs, with different, controllable characteristics.
In general, any combination of mapped behavioral deviation patterns can be applied to $M_0$, creating a set of $n$ models for $M^S$, with $n$ the number of such combinations. Subsequently, the same can be done for mapped recording error patterns, on each $M^S$, resulting in a set of $n \cdot m$ models for $M^L$, with $m$ the number of combinations of recording errors. Lastly, with $k$ different simulation parameters, $n\cdot m \cdot k$ logs can be generated from $M^0$, each with their own distinct characteristics with regard to included behavioral deviations, recording errors, and stochastic. Fig.~\ref{fig:framework_extension} shows this extended usage of the framework.}

\tent{In this section, we provide three instances of using the framework as described by Fig.~\ref{fig:framework_extension} to create synthetic datasets of processes following the description of Sec.~\ref{sec:framework_design}. For each, we describe the complete process, \ie from designing a base model, to selecting and mapping deviation patterns, to choosing simulation parameters to generate several event logs. We reflect on their properties and compare qualitatively the characteristics of the datasets with regard to the capabilities of other synthetic data generation methods from Sec.~\ref{sec:relwork_synth}.}

\tent{All models and corresponding generated logs with the applied patterns are available at \href{https://gitlab.com/dominiquesommers/mira/-/tree/main/mira/simulation}{gitlab.com/dominiquesommers/mira/-/tree/main/mira/simulation}.}

\subsection{\tent{Package delivery process}}\label{sec:packageprocess}
The first instance follows from the running example of the package delivery process (\cf Fig.~\ref{fig:M_package}) used throughout the previous sections. This synthetically designed process includes several modeling patterns as desribed in Sec.~\ref{sec:chars}. Let us say that various recording mechanisms are incorporated to record the activities, \eg orders are registered in an information system, and the activities in which human actors are involved are logged by them manually scanning the packages' barcodes. For the delivery at a depot, the corresponding depot is registered by the courier.

Note that this process is designed in a way such that it could indicate a real-life process. We deem this an important property as to reflect realistic characteristics.

For this demonstration, we formalized this process as a t-PNID which serves as the base process model $M_0$, and describes the expected behavior of the package delivery process.

\begin{figure}[tb]
\centering
\hspace*{-1.8cm}
\includegraphics[scale=0.65]{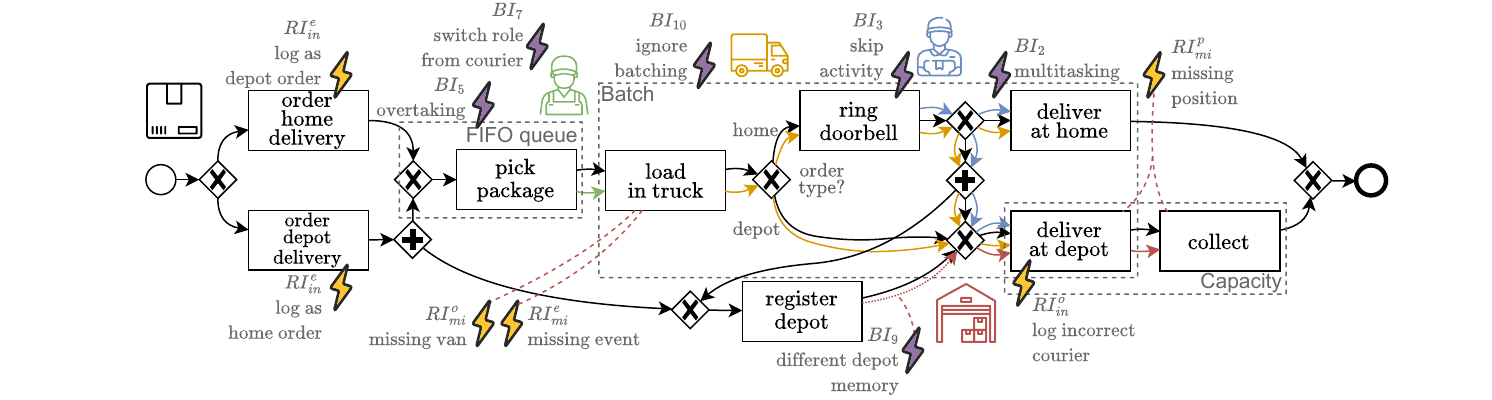}
\hspace*{1.8cm}
\vspace*{-0.3cm}
    \caption{Abstract and simplified base process model for the package delivery process from Fig.~\ref{fig:M_package}, with additionally annotated the added behavioral outliers (purple thunderbolts) and recording errors (yellow thunderbolts).}
    \label{fig:M_package_dev}
\end{figure}

\tent{\emph{Behavioral deviations:} The following behavioral deviation patterns with corresponding mappings are applied to $M_0$ as illustrated by the purple deviations in Fig.~\ref{fig:M_package_dev}:
\begin{itemize}
    \item[$BI_5$] \emph{Overtaking} in the FIFO queue for \emph{picking packages};
    \item[$BI_7$] \emph{Switching roles} from a \emph{courier} to that of a \emph{warehouse employee};
    \item[$BI_{10}$] \emph{Batching is ignored}, \emph{leaving with a delivery van} before it was fully loaded;
    \item[$BI_3$] \emph{Skipping the activity} of \emph{ringing}, modeling behavior where \eg the door was already opened upon arrival;
    \item[$BI_9$] \emph{Different resource memory} where the package is \emph{delivered} to a different \emph{depot} than where it is \emph{registered};
    \item[$BI_2$] \emph{Multitasking} of \emph{couriers} during the \emph{delivery} of multiple packages, modeling interruption of a delivery.
\end{itemize}}

\tent{For this dataset, we choose to isolate the patterns, not combining them within a single simulated event log. We therefore create six versions for $M^S$, each applying a single behavioral deviation pattern model transformation from the list above. Additionally, we add a version for $M^S$, denoted as $M^S_\emptyset$ without behavioral deviations.}

\tent{For recording errors, the following patterns with corresponding mappings are applied to $M^S_\emptyset$, illustrated by the yellow deviations in Fig.~\ref{fig:M_package_dev}:
\begin{itemize}
    \item[$RI_{in}^e (1)$] \emph{Incorrect event}, recording an \emph{order for depot delivery} when it was intended for \emph{home delivery};
    \item[$RI_{in}^e (2)$] \emph{Incorrect event}, vice versa, \ie recording an \emph{order for home delivery} when it was intended for \emph{depot delivery};
    \item[$RI_{mi}^e$] \emph{Missing event} for the activity of \emph{loading} a package in a truck;
    \item[$RI_{mi}^o$] \emph{Missing object} of the involved \emph{van} for \emph{loading}, \eg due to a temporary connection failure of a recording device;
    \item[$RI_{in}^o$] \emph{Incorrect object} of the \emph{courier} when \emph{ringing}, \eg due to not logging out by the courier on the previous shift;
    \item[$RI_{mi}^p$] \emph{Missing positions} for the recording of the \emph{delivery} and the \emph{collection} at a depot, \eg due to coarse timestamp logging.
\end{itemize}}

\begin{figure}[tb]
\centering
\includegraphics[scale=0.65]{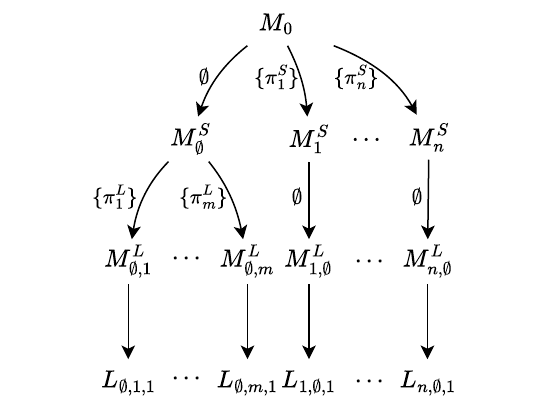}
    \caption{\tent{Framework as applied to the synthetic package delivery process.}}
    \label{fig:framework_applied1}
\end{figure}

\tent{For the other versions of $M^S$, no recording error patterns are applied. This results in a set of twelve models representing $M^L$, six of which with a distinct behavioral deviation pattern and six of which with a distinct recording error pattern. For this process, we select a single set of simulation parameters that ensure that the patterns are invoked in the simulation of a small number of packages (two) in each of these twelve models, as illustrated in Fig.~\ref{fig:framework_applied1}. The twelve generated event logs each contain the behavior for handling the two packages with the corresponding issue, be it either a behavioral deviation or a recording error.}\\

\tent{Note that it is actually not inherently infeasible to generate the same (or similar) event logs as we described here, using the existing synthetic process data generation methods from Sec.~\ref{sec:relwork_synth}. For each behavioral deviation pattern, a separate process model can be constructed, and similar recording errors can be introduced, either with careful settings regarding the noise injection, or manually. However, we argue that creating the dataset following our framework provides a methodological and structured approach to its design, where the incorporated issues stem from the literature on real-life process data. As the patterns for both behavioral deviations and recording errors are designed to be additive in behavior, the expected behavior of the process does not change. Next, we demonstrate using our framework for creating another dataset with deviation patterns that are not directly supported by the existing methods Sec.~\ref{sec:relwork_synth}.}

\subsection{\tent{Energy contract process}}
\begin{figure}[tb]
\centering
\hspace*{-1.85cm}
\includegraphics[scale=0.65]{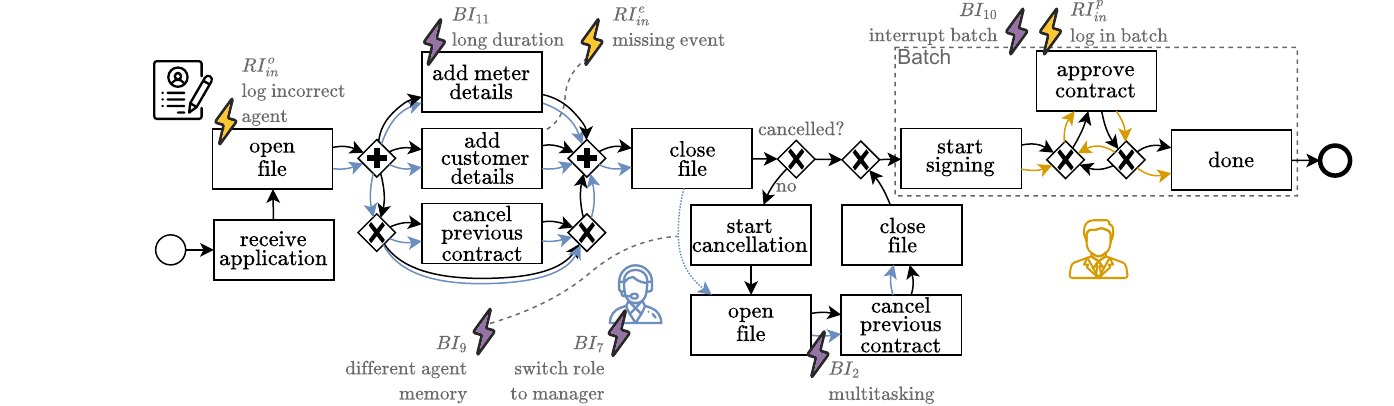}
\hspace*{1.85cm}
\vspace*{-0.3cm}
    \caption{\tent{Abstract and simplified base process model for the energy contract process from, with additionally annotated the behavioral outliers (purple thunderbolts) and recording errors (yellow thunderbolts) that are applied.}}
    \label{fig:M_energy_dev}
\end{figure}

\tent{For our second dataset, we design again a synthetic process originating from a process mining course given at our university. The process considers the handling of energy contracts and is illustrated in Fig.~\ref{fig:M_energy_dev}. It is in terms of structure and involved objects less complicated than the package delivery process, however, it includes some additional properties.}

\tent{Customers can apply for a new energy contract online, after which a few pre-checks are performed by the company. After the pre-checks when \emph{receiving} the application, the employees in the back office \emph{add the customer details} to the database. They also \emph{add the meter details}. As a service, the previous contract of the customer is \emph{canceled}. Finally, a manager does a final check and \emph{approves} the contract. We see three object types in this process: one for \emph{applications} (black), one for the \emph{call agents} (blue) in the back office, and one for \emph{managers} (orange).}

\tent{In the subprocess of adding the details, which are executed in parallel by a single agent, the call agent is expected to work uninterruptedly on an application, which starts and ends with opening and closing the file. Cancellation of the previous contracts is either done during this subprocess or in a later stage, for which the agent reopens the file, resulting in activities with duplicate labels in the base process model.}

\tent{Managers are expected to approve contracts in batches. Batching is defined slightly differently here than in the package delivery process, where the delivery can only be initiated after a batch is complete, \ie a van is fully loaded. Here, a manager can initiate the handling of contracts at any point in time (\emph{start signing}) and they finish the approvals of each contract in the taken batch before resuming with newly arrived contracts.}

\tent{The model from Fig.~\ref{fig:M_energy_dev} is again formalized as a t-PNID and serves as the base process model $M_0$, describing the expected behavior of the energy contract process.}

\tent{We add the following behavioral deviation patterns with corresponding mappings to $M_0$ as depicted by the purple deviations in Fig.~\ref{fig:M_energy_dev}:
\begin{itemize}
    \item[$BI_7$] \emph{Switching roles} where an \emph{agent} temporarily takes over the role of the \emph{manager};
    \item[$BI_9$] \emph{Different resource memory} where a different \emph{agent} does the second phase \emph{cancelation};
    \item[$BI_{10}$] \emph{Ignore batching} of \emph{approving contracts}, interrupting the handling of a batch and contracts are added to it before the batch finishes;
    \item[$BI_2$] \emph{Multitasking} of \emph{agents} in the second phase \emph{cancellation}, \eg in case where the first phase cancellation requires more attention than the second;
    \item[$BI_{11}$] \emph{Long duration} of \emph{adding meter details} taking exceptionally long with regard to the usual processing times.
\end{itemize}}

\tent{We combine each deviation pattern in one model, creating a single version for $M^S$, to which we add the following recording error patterns with corresponding mappings, to create a single version for $M^L$, as depicted by the yellow deviations in Fig.~\ref{fig:M_energy_dev}:
\begin{itemize}
    \item[$RI_{mi}^e$] \emph{Missing event} for the activity of \emph{adding customer details};
    \item[$RI_{in}^o$] \emph{Incorrect object} of the involved \emph{agent} in the first phase of handling applications, \eg due to not logging out by the agent on the previous shift;
    \item[$RI_{in}^o$] \emph{Incorrect object} where the activity of \emph{cancel previous contract} in the first phase is accompanied with the data label of \emph{false}, possibly resulting in a second cancellation;
    \item[$RI_{in}^p$] \emph{Incorrect position} due to batch logging of \emph{approving contracts}, causing a batch of contracts to be approved seemingly at the same time.
\end{itemize}}

\tent{$M^L$ contains each deviation pattern described above, which, with simulation parameters, is ready for simulating corresponding event logs. We select simulation parameters such that the frequencies of the occurrences of issues are low, and set the a duration distribution for deviation $BI_{11}$ such that it deviation from the existing activity for adding meter details. By simulating the process for a large number of contract applications, we assume that by the law of large numbers, each issue will appear in the generated logs with approximately the expected frequencies. Note that guided simulation approach like PURPLE~\cite{burattin2022purpose} (\cf Sec.~\ref{sec:relwork_synth}) can complement the simulation model in our framework to ensure characteristics of the event log regarding the frequencies of included issues.}\\

\tent{Some of the incorporated issues are applied to activities with duplicate labels in the base model. Differentiating between such activities, and only applying certain issues on one of them, is infeasible without the connection of the simulated recorded events and the process model(s). Therefore, noise injection by random imputations performed on a simulated event log without deviations does not support these issues. Also due to the duplicate labels, the base model is not discoverable by process discovery methods like the Alpha Miner~\cite{DBLP:journals/tkde/AalstWM04} or the Inductive Miner~\cite{DBLP:conf/bpm/LeemansFA13}, event for simulated event logs without any deviations. 
Hence, we need the corresponding models, \ie $M_0$, $M^S$, and $M^L$, for the assessment with a dataset like the one created for this energy contract process.}

\subsection{\tent{Real-life assembly process}}
\tent{The third dataset revolves around the real-life manufacturing process from our partner company Omron that we referred to in Sec.~\ref{sec:introduction}. This demonstrates the usage of the framework in a different way, for which the base process model already exists in the form of a normative model and we anticipate issues in the process, both for behavioral deviations and recording errors. Recall from Sec.~\ref{sec:introduction} that discrepancies between the model and event data of this process, \eg exposed through alignments, can be explained in various ways with no way of verifying the explanation. However, if we add patterns for the different explanations to the base model, we can generate the corresponding scenarios synthetically.}

\begin{figure}[tb]
\centering
\hspace*{-1.6cm}
\includegraphics[scale=0.65]{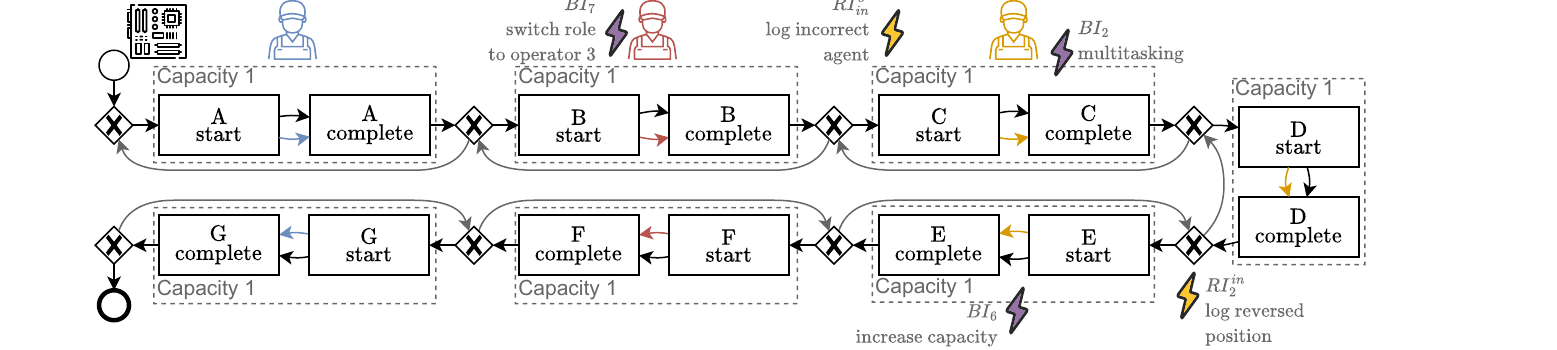}
\hspace*{1.6cm}
\vspace*{-0.3cm}
    \caption{\tent{Abstract simplified normative process model for a real-life assembly process.}}
    \label{fig:M_omron_dev}
\end{figure}

\tent{Fig.~\ref{fig:M_omron_dev} shows an abstract and simplified model of the manufacturing process. This model is created in collaboration with the process owners and acts as the base model $M_0$ in the framework. \emph{Products} (black) are assembled in seven sequential stages, \ie stage A to stage G, each containing a predefined sequence of activities. Three operators are responsible for their own stages, \ie stages A and G for operator $1$ (blue), stages B and F for operator $2$ (red), and stages C, D, and E for operator $3$ (orange). This is based on the physical layout of the stages in the factory, which is roughly resembled by the layout of $M_0$ in Fig.~\ref{fig:M_omron_dev}. A stage can only process one product at the same time, as depicted by the capacities. In between stages, there is always the option for a product to be put back to any of the previous stages, through (a series of) reversions.}

\tent{Activity executions in this process are recorded by manually scanning the corresponding product at the start and completion of an activity. The involved operator object is added automatically and is decided by the hardcoded responsibilities for activities.}

\tent{For completeness, we consider all seven stages, however for conciseness, we add patterns to half of the process representing the explanations for deviations from Sec.~\ref{sec:introduction}. Note that it can be trivially extended to the other stages and operators as well. We add the following three behavioral deviation patterns to $M_0$ to create $M^S$ and subsequently two recording error patterns to $M^S$ to create $M^L$ with corresponding mappings:
\begin{itemize}
    \item[$BI_6$] \emph{Increase capacity} of \emph{stage E}, allowing for more than one product to be handled simultaneously;
    \item[$BI_7$] \emph{Switching roles} from \emph{operator 2} to \emph{operator 3}, \eg to help out when fully occupied;
    \item[$BI_2$] \emph{Multitasking} of \emph{operator 2}, handling multiple products in various stages simultaneously;
    \item[$RI_{in}^o$] \emph{Incorrect object} of the involved \emph{operator} which is inherently accompanied by the combination of switching roles, due to the working of the recording mechanism. Recall that to the involved operator is added automatically, hence activities in \emph{stages C to E} involving operator 2 are automatically logged on operator 3.
    \item[$RI_{in}^p$] \emph{Incorrect position} for the recording of activities in \emph{stages D and E}, caused by manual logging of these activities which is sensitive to incoherent positioning of the recorded events as opposed to when they were actually executed.
\end{itemize}}

\tent{For the dataset described here, we simulate $M^L$ for this process using standard simulation parameters to include the `what-if' scenarios caused by the included issues. Note that one can add contextual information and data to the conditions of stochastics encoded in the simulation model to increase the authenticity of the occurrences of the patterns, \eg using the current state to increase the probability of switching roles when operator 3 is busy. Furthermore, stochastic information from the previously recorded data can be used to ultimately create an event log that is realistic with regard to the underlying real-life process.}\\

\tent{The patterns of the issues applied to the base process model are all behaviorally influencing each other, which is not something that is supported by the existing methods from Sec.~\ref{sec:relwork_synth}.}

\section{Evaluating the interpretability of (Relaxed) multi-object alignments}\label{sec:alignments}
As mentioned in Sec.~\ref{sec:related_work}, alignments are used in conformance checking to expose deviations and provide possible explanations for discrepancies between a process model and a recorded event log. Here, we demonstrate our ground truth assessment approach for qualitative validation and evaluation of three alignment techniques on the synthetically designed package delivery process and the corresponding generated event logs as described in Sec.~\ref{sec:packageprocess}. This provides a concrete example of the usage of the framework for an assessment of process mining techniques, following the description from Sec.~\ref{sec:usage}.

\subsection{Experimental setup}

\begin{figure*}[tb]
\centering
\includegraphics[scale=0.75]{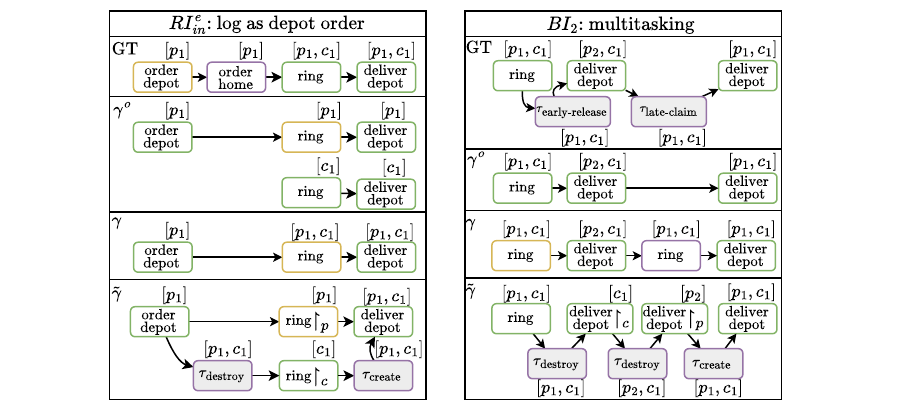}
\vspace*{-0.05cm}
\caption{Ground truth (GT) and computed ($\al^o,\al$, and $\relal$) sub-alignment results for recording error pattern $RI_{in}^e$ and behavioral deviation pattern $BI_2$. Log, model, and synchronous moves are colored yellow, purple, and green respectively. For clarity, truck and depot objects are omitted.}
    \label{fig:result_alignments}
\end{figure*}

The assessment question (AQ) we aim to answer is threefold:\\
\\
\emph{AQ1: Are behavioral outliers and recording errors detected correctly? 
\\
AQ2: How well do the generated alignments reflect the ground truth explanations?\\
\tent{AQ3: How do the behavioral outliers and recording errors affect the computation times of the alignment methods?}}\\

With six behavioral deviation patterns and six recording error patterns, the dataset created from the synthetic package delivery process contains twelve versions of $M^L$ and twelve corresponding event logs, each containing a single issue. This isolates the deviation patterns with each event log handling two package orders using the same delivery van object. The ground truth oracle provides explanations by knowledge of which transitions have fired in the simulation to produce the events in the logs.

$M_0$ and the event logs serve as input for the three alignment methods we consider in this assessment. These are (1) traditional \emph{per-object alignments} ($\al^o$), operating on each object in isolation, both in event log and model, ignoring any interaction between objects~\cite{carmona2018conformance}, (2) \emph{systemic} alignments ($\al$), aligning the event log in its entirety to the process model taking into account violations~\cite{sommers2022aligning}, and (3) \emph{relaxed systemic} alignments ($\relal$), extending their regular counterpart by allowing for relaxations of objects' interactions both in the log and model~\cite{sommers2024conformance}.
 
Each method takes an event log being a partially ordered set and a normative model as input. The log used for the alignments is a projection of the simulated logs with some types of deviating events (like skipped steps or missing events) removed from it.
The produced alignment contains synchronous moves (on which the log and model agree), log moves (indicating that an activity from the log cannot be mimicked in the model) and model moves (indicating which events required by the model are missing in the log).

\begin{figure}[tb]
\centering
\captionsetup{width=\linewidth}
\captionof{table}{\tent{Experiment results showing the objects responsible for the deviation and those affected by it (in subscript) for each pattern, for the ground truth (GT) and detected by alignment methods: per-object ($\al^o$), systemic ($\al$), and relaxed systemic ($\relal$).}}\label{tab:results}
\hspace*{-.8cm}
\includegraphics[scale=0.85]{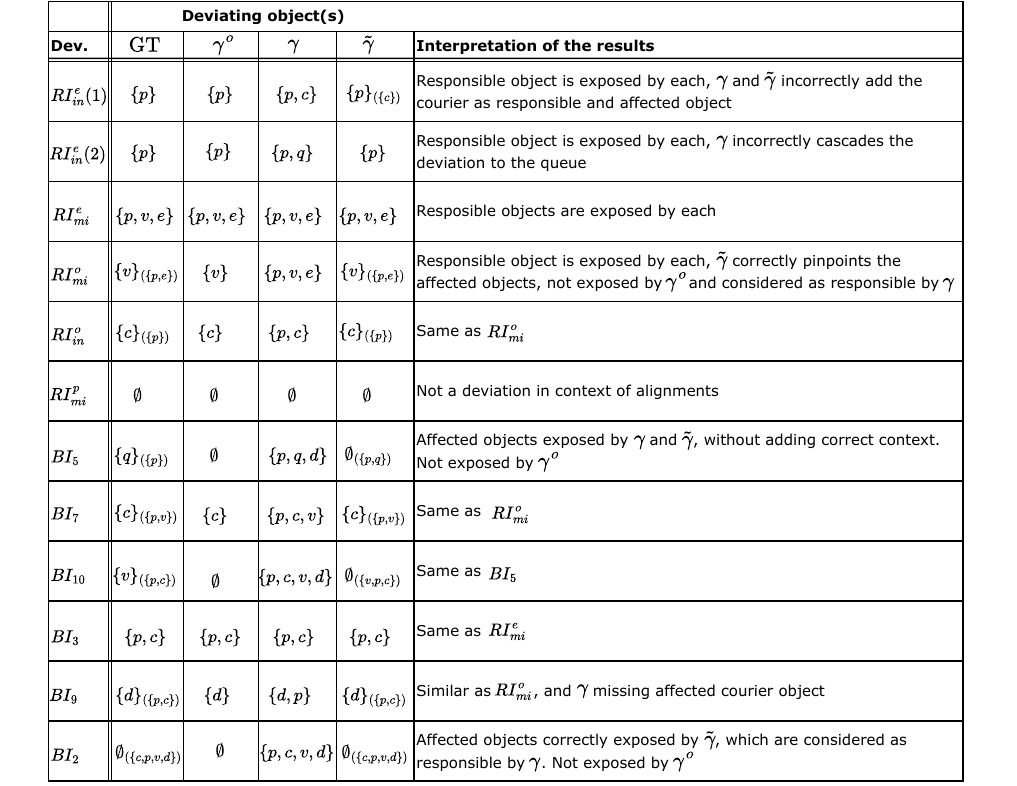}
\hspace*{.8cm}
\end{figure}

\subsection{Results and discussion}
The detected deviations for each log are summarized in Tab.~\ref{tab:results}. Each row in the table corresponds to a log. The Dev. column indicates the deviation used in the corresponding log. The place of the deviations in the model is shown in Fig.~\ref{fig:M_package_dev}. The cells of the table specify the set of objects included in log and model moves of the corresponding alignments, and in addition the set of affected objects in brackets (in the subscript) where applicable. The right-most column summarizes the interpretations of the results.

For recording errors, each deviation was detected by each method, however, there is a difference in the involved objects when the deviation occurs on object level rather than event level. For the single-object alignment $\al^o$, we see only the deviation's responsible object, for the system alignment $\al$, we see all objects involved in the corresponding activity, and for the relaxed systemic alignment $\relal$, we see a combination of the two, correctly separating responsible objects from affected objects. The last recording error ($RI_{mi}^p$) is not considered a deviation in the context of alignments.
From the computed alignments for $RI_{in}^e(1)$ (\cf Fig.~\ref{fig:result_alignments}), we see that none of the methods resolve the deviation correctly, as all conclude that the \emph{ring} event is incorrectly logged instead of the \emph{order depot} event. The sub-alignments clearly show how each method distinguishes between the involved objects.

For behavioral outliers, the two deviations $BI_7$ and $BI_3$ occur on the level of isolated activity or object and are detected similarly as the recording errors discussed above. The others are not detected by single-object alignments, as they correspond to deviations in objects' interactions, which are not considered. Except for $BI_9$, where the executed behavior of both depot objects is incomplete according to the model. 

Systemic alignments $\al$ expose all deviations, again noting all involved objects without indication of the object(s) responsible for the deviation. This distinction is made by relaxed alignments $\relal$ for deviations $BI_7$ and $BI_3$, while for the remaining deviations, the involved objects are detected without log and (labeled) model moves. 

From the alignments computed for $BI_2$ (\cf Fig.~\ref{fig:result_alignments}), we can see the consequence of ignoring correlations of objects, as the single-object alignments $\al^o$ show no deviations. This is resolved in the systemic alignment $\al$. However, its explanation differs from the deviation pattern used to generate the log, contrary to the relaxed systemic alignment $\relal$, which includes (relaxed) synchronous moves for each recorded event and alterings of correlations. Note that these alignments still allow for variance in their interpretation, \eg the separated synchronous moves for deliver depot and the non-matching silent model moves.

\begin{figure}[tb]
\centering
\hspace*{-1.1cm}
\includegraphics[scale=0.5]{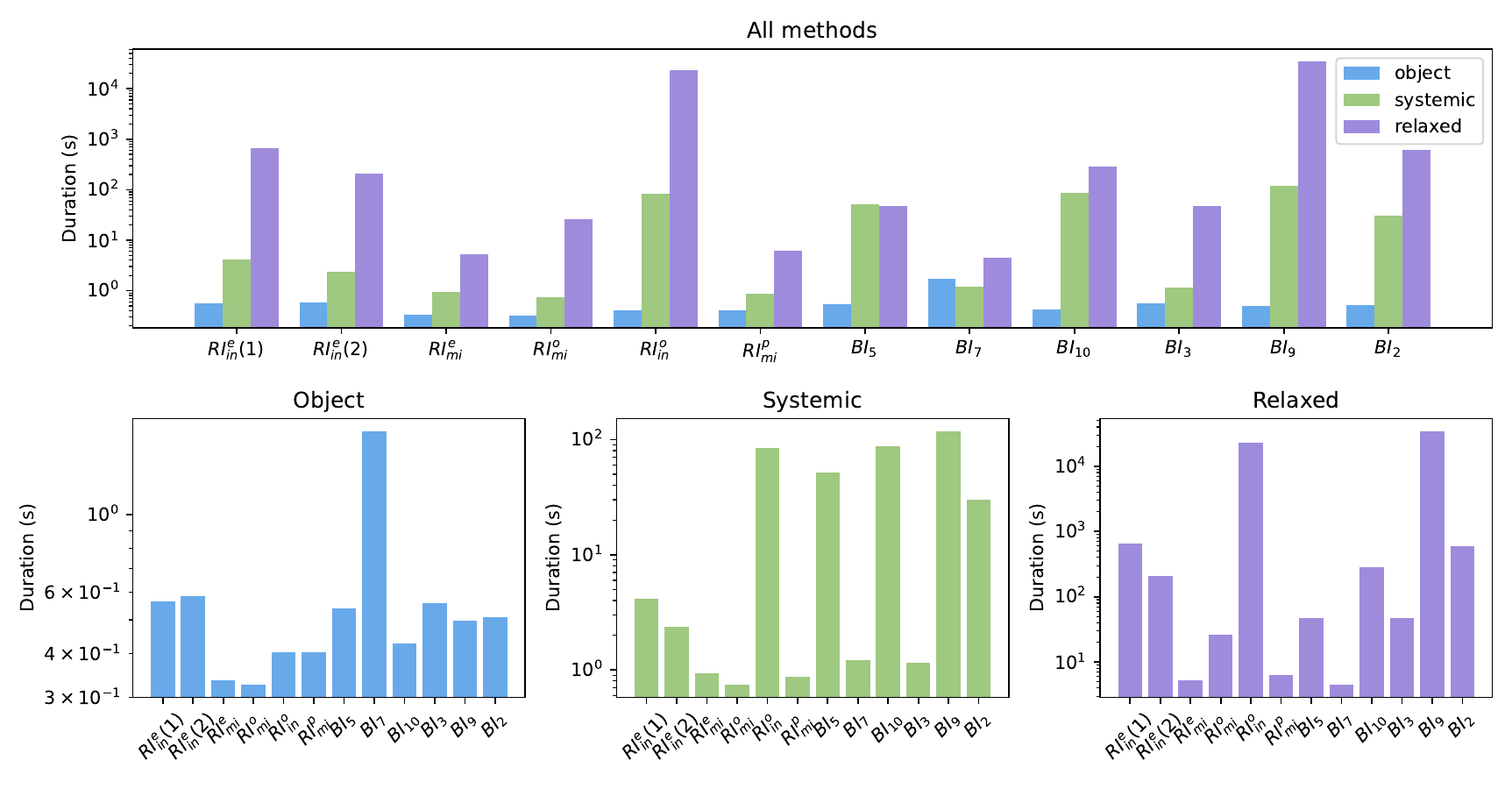}
\hspace*{1.1cm}
\vspace*{-0.05cm}
\caption{\tent{Computation times of each method for every pattern of recording errors and behavioral deviations.}}
    \label{fig:result_durations}
\end{figure}
\tent{Fig.~\ref{fig:result_durations} shows the durations of computing the alignment for the three methods on the log for each pattern of recording errors and behavioral deviation. Looking at the graph in the top row, we can compare the computation times for the three methods for each pattern. The results distinguish three `patterns` regarding the increase in computation time from the less to the more demanding techniques: (1) For each pattern except for $BI_5$, $BI_{10}$, and $BI_7$, we see an exponential increase for the three methods. (2) For $BI_5$ and $BI_{10}$, the computation time increases similarly from object alignments to systemic alignments, but there is no significant increase to relaxed alignments. (3) For $BI_7$, there is no significant increase in computation time between any of the three methods.}

\tent{The graphs on the bottom row isolate the data for the three different alignment techniques, showing that the computation times for object alignments are similar across the patterns, with one exception for switching roles ($BI_7$). Systemic alignments show two patterns, where the computation is either fast or significantly slower. The graph shows more variance for relaxed alignments, with a reasonably constant increase in computation time from systemic alignments, apart from the patterns $BI_5$, $BI_7$, $BI_{10}$, which is in line with the patterns from the graph in the top row.}

In general, to answer AQ1 and AQ2, we see that single-object alignments can pinpoint deviations on the level of the responsible object, but only if they occur within the workflow of the particular object. Systemic alignments are robust in detecting the deviations but fail to provide details for the responsible objects. Relaxed systemic alignments do both, however, not always provide the correct explanation. \tent{We see a clear trade-off between interpretable deviations and computation times. To answer AQ3, there are certain patterns, especially for behavioral deviations, where more interpretability can be achieved without significantly giving in on the computation times, \eg for $BI_7$, $BI_5$, and $BI_{10}$.}

This helps to outline the directions for future work in this area. Having the ground truth knowledge provided by our framework was indispensable for conducting this assessment.

\section{Conclusion and Future Work}\label{sec:conclusion}
Where real-life process data provides for valuable assessments of process mining techniques used in realistic scenarios, the absence of an omniscient stakeholder and specific process characteristics may limit the validity and generalizability of the assessment results. The use of synthetic data allows extending the scale of experimentation and creating an oracle generating the ground truth to be used in assessment.

The approach proposed in this paper makes use of deviations in terms of behavioral outliers and recording errors known from the literature and captures them as modeling patterns and corresponding model transformations. Starting from a base model, we generate separate models for an executed process and for recorded process behavior, as well as a synthetically generated yet realistically imperfect event log. This approach is limited by the assumption that the deviations can be modeled using the same formalism as the initial model, and by the dependency on the simulation method and parameters regarding the frequencies of the incorporated deviations, potentially affecting the realism of the dataset.
The list of deviations described (and implemented) in this work is inherently incomplete and is designed to be extended depending on the domain. \Eg it naturally extends to the stochastic and temporal aspects of the process, which can either be modeled in patterns or the simulation module directly.

\tent{We described the usage of the framework for creating datasets consisting of multiple event logs from manually designed base models for three synthetic processes. We showed that the characteristics represented by the datasets regarding recording errors and behavioral outliers are not supported by existing synthetic process data generation methods.}

Through a demonstration on one of the created datasets, we show that this approach provides validation and qualitative insights into the \tent{strengths, weaknesses, and computation times} of different alignment techniques for the detection and explanation of deviations in modeled and recorded behavior. The results can be differentiated per deviation type, \ie recording errors and behavioral outliers, as well as per deviation itself, and tailored towards domain-specific parts of the process.

Whereas in this work, we focused on the \emph{qualitative} aspect of the \emph{evaluation}, this approach trivially extends to the \emph{quantitative} aspect, by providing a broader dataset with varying frequencies of deviating behavior and analyzing the corresponding statistics of the results. Subsequently, it allows for a \emph{reliability}, \emph{robustness}, \emph{performance}, and \emph{usability} assessment, quantitatively as well as qualitatively, providing results with regard to characteristics of the process, \ie in terms of domain-specific (expected) deviating behavior.

\section*{Declarations}
\subsection*{Ethical approval}
Not applicable.

\subsection*{Funding}
This work is done within the project ``Certification of production process quality through Artificial Intelligence (CERTIF-AI)'', funded by NWO (project number: \href{https://www.nwo.nl/projecten/17998}{17998}).

\subsection*{Availability of data and materials}
All datasets generated and used for analysis in this study are available at \href{https://gitlab.com/dominiquesommers/mira/-/tree/main/mira/simulation}{gitlab.com/dominiquesommers/mira/-/tree/main/mira/simulation}.

\bibliography{main}


\begin{thebibliography}{34}
\ifx \bisbn   \undefined \def \bisbn  #1{ISBN #1}\fi
\ifx \binits  \undefined \def \binits#1{#1}\fi
\ifx \bauthor  \undefined \def \bauthor#1{#1}\fi
\ifx \batitle  \undefined \def \batitle#1{#1}\fi
\ifx \bjtitle  \undefined \def \bjtitle#1{#1}\fi
\ifx \bvolume  \undefined \def \bvolume#1{\textbf{#1}}\fi
\ifx \byear  \undefined \def \byear#1{#1}\fi
\ifx \bissue  \undefined \def \bissue#1{#1}\fi
\ifx \bfpage  \undefined \def \bfpage#1{#1}\fi
\ifx \blpage  \undefined \def \blpage #1{#1}\fi
\ifx \burl  \undefined \def \burl#1{\textsf{#1}}\fi
\ifx \doiurl  \undefined \def \doiurl#1{\url{https://doi.org/#1}}\fi
\ifx \betal  \undefined \def \betal{\textit{et al.}}\fi
\ifx \binstitute  \undefined \def \binstitute#1{#1}\fi
\ifx \binstitutionaled  \undefined \def \binstitutionaled#1{#1}\fi
\ifx \bctitle  \undefined \def \bctitle#1{#1}\fi
\ifx \beditor  \undefined \def \beditor#1{#1}\fi
\ifx \bpublisher  \undefined \def \bpublisher#1{#1}\fi
\ifx \bbtitle  \undefined \def \bbtitle#1{#1}\fi
\ifx \bedition  \undefined \def \bedition#1{#1}\fi
\ifx \bseriesno  \undefined \def \bseriesno#1{#1}\fi
\ifx \blocation  \undefined \def \blocation#1{#1}\fi
\ifx \bsertitle  \undefined \def \bsertitle#1{#1}\fi
\ifx \bsnm \undefined \def \bsnm#1{#1}\fi
\ifx \bsuffix \undefined \def \bsuffix#1{#1}\fi
\ifx \bparticle \undefined \def \bparticle#1{#1}\fi
\ifx \barticle \undefined \def \barticle#1{#1}\fi
\bibcommenthead
\ifx \bconfdate \undefined \def \bconfdate #1{#1}\fi
\ifx \botherref \undefined \def \botherref #1{#1}\fi
\ifx \url \undefined \def \url#1{\textsf{#1}}\fi
\ifx \bchapter \undefined \def \bchapter#1{#1}\fi
\ifx \bbook \undefined \def \bbook#1{#1}\fi
\ifx \bcomment \undefined \def \bcomment#1{#1}\fi
\ifx \oauthor \undefined \def \oauthor#1{#1}\fi
\ifx \citeauthoryear \undefined \def \citeauthoryear#1{#1}\fi
\ifx \endbibitem  \undefined \def \endbibitem {}\fi
\ifx \bconflocation  \undefined \def \bconflocation#1{#1}\fi
\ifx \arxivurl  \undefined \def \arxivurl#1{\textsf{#1}}\fi
\csname PreBibitemsHook\endcsname

\bibitem[\protect\citeauthoryear{Rozinat et~al.}{2008}]{rozinat2008}
\begin{bchapter}
\bauthor{\bsnm{Rozinat}, \binits{A.}},
\bauthor{\bsnm{Medeiros}, \binits{A.K.A.}},
\bauthor{\bsnm{G{\"u}nther}, \binits{C.W.}},
\bauthor{\bsnm{Weijters}, \binits{A.J.M.M.}},
\bauthor{\bsnm{Aalst}, \binits{W.M.P.}}:
\bctitle{The need for a process mining evaluation framework in research and practice}.
In: \beditor{\bsnm{Hofstede}, \binits{A.}},
\beditor{\bsnm{Benatallah}, \binits{B.}},
\beditor{\bsnm{Paik}, \binits{H.-Y.}} (eds.)
\bbtitle{Business Process Management Workshops},
pp. \bfpage{84}--\blpage{89}.
\bpublisher{Springer},
\blocation{Berlin, Heidelberg}
(\byear{2008})
\end{bchapter}
\endbibitem

\bibitem[\protect\citeauthoryear{Jouck and Depaire}{2019}]{jouck2019generating}
\begin{barticle}
\bauthor{\bsnm{Jouck}, \binits{T.}},
\bauthor{\bsnm{Depaire}, \binits{B.}}:
\batitle{Generating artificial data for empirical analysis of control-flow discovery algorithms: A process tree and log generator}.
\bjtitle{Business \& Information Systems Engineering}
\bvolume{61},
\bfpage{695}--\blpage{712}
(\byear{2019})
\end{barticle}
\endbibitem

\bibitem[\protect\citeauthoryear{Burattin et~al.}{2022}]{burattin2022purpose}
\begin{bchapter}
\bauthor{\bsnm{Burattin}, \binits{A.}},
\bauthor{\bsnm{Re}, \binits{B.}},
\bauthor{\bsnm{Rossi}, \binits{L.}},
\bauthor{\bsnm{Tiezzi}, \binits{F.}}:
\bctitle{A purpose-guided log generation framework}.
In: \bbtitle{International Conference on Business Process Management},
pp. \bfpage{181}--\blpage{198}.
\bpublisher{Springer},
\blocation{Cham}
(\byear{2022})
\end{bchapter}
\endbibitem

\bibitem[\protect\citeauthoryear{Ko et~al.}{2020}]{ko2020air}
\begin{bchapter}
\bauthor{\bsnm{Ko}, \binits{J.}},
\bauthor{\bsnm{Lee}, \binits{J.}},
\bauthor{\bsnm{Comuzzi}, \binits{M.}}:
\bctitle{{AIR-BAGEL}: An interactive root cause-based anomaly generator for event logs.}
In: \bbtitle{ICPM Doctoral Consortium/Tools},
vol. \bseriesno{2703},
pp. \bfpage{35}--\blpage{38}.
\bpublisher{CEUR Workshop Proceedings},
\blocation{Aachen}
(\byear{2020})
\end{bchapter}
\endbibitem

\bibitem[\protect\citeauthoryear{Sommers et~al.}{2024}]{sommers2024assessing}
\begin{bchapter}
\bauthor{\bsnm{Sommers}, \binits{D.}},
\bauthor{\bsnm{Sidorova}, \binits{N.}},
\bauthor{\bsnm{Dongen}, \binits{B.F.}}:
\bctitle{Assessing process mining techniques: a ground truth approach}.
In: \bbtitle{2024 6th International Conference on Process Mining (ICPM)},
pp. \bfpage{41}--\blpage{48}
(\byear{2024}).
\doiurl{10.1109/ICPM63005.2024.10680619}
\end{bchapter}
\endbibitem

\bibitem[\protect\citeauthoryear{Dees et~al.}{2017}]{dees2017}
\begin{bchapter}
\bauthor{\bsnm{Dees}, \binits{M.}},
\bauthor{\bsnm{Leoni}, \binits{M.}},
\bauthor{\bsnm{Mannhardt}, \binits{F.}}:
\bctitle{Enhancing process models to improve business performance: A methodology and case studies}.
In: \beditor{\bsnm{Panetto}, \binits{H.}},
\beditor{\bsnm{Debruyne}, \binits{C.}},
\beditor{\bsnm{Gaaloul}, \binits{W.}},
\beditor{\bsnm{Papazoglou}, \binits{M.}},
\beditor{\bsnm{Paschke}, \binits{A.}},
\beditor{\bsnm{Ardagna}, \binits{C.A.}},
\beditor{\bsnm{Meersman}, \binits{R.}} (eds.)
\bbtitle{On the Move to Meaningful Internet Systems. OTM 2017 Conferences},
pp. \bfpage{232}--\blpage{251}.
\bpublisher{Springer},
\blocation{Cham}
(\byear{2017})
\end{bchapter}
\endbibitem

\bibitem[\protect\citeauthoryear{Bose et~al.}{2013}]{bose2013wanna}
\begin{bchapter}
\bauthor{\bsnm{Bose}, \binits{R.P.J.C.}},
\bauthor{\bsnm{Mans}, \binits{R.S.}},
\bauthor{\bsnm{Aalst}, \binits{W.M.P.}}:
\bctitle{Wanna improve process mining results?}
In: \bbtitle{2013 IEEE Symposium on Computational Intelligence and Data Mining (CIDM)},
pp. \bfpage{127}--\blpage{134}
(\byear{2013}).
\doiurl{10.1109/CIDM.2013.6597227}
\end{bchapter}
\endbibitem

\bibitem[\protect\citeauthoryear{K{\"a}ppel et~al.}{2021}]{kappel2021}
\begin{bchapter}
\bauthor{\bsnm{K{\"a}ppel}, \binits{M.}},
\bauthor{\bsnm{Jablonski}, \binits{S.}},
\bauthor{\bsnm{Sch{\"o}nig}, \binits{S.}}:
\bctitle{Evaluating predictive business process monitoring approaches on small event logs}.
In: \beditor{\bsnm{Paiva}, \binits{A.C.R.}},
\beditor{\bsnm{Cavalli}, \binits{A.R.}},
\beditor{\bsnm{Ventura~Martins}, \binits{P.}},
\beditor{\bsnm{P{\'e}rez-Castillo}, \binits{R.}} (eds.)
\bbtitle{Quality of Information and Communications Technology},
pp. \bfpage{167}--\blpage{182}.
\bpublisher{Springer},
\blocation{Cham}
(\byear{2021})
\end{bchapter}
\endbibitem

\bibitem[\protect\citeauthoryear{G{\"u}nther}{2009}]{gunther2009process}
\begin{botherref}
\oauthor{\bsnm{G{\"u}nther}, \binits{C.W.}}:
Process mining in flexible environments.
Phd thesis,
Eindhoven University of Technology, Industrial Engineering and Innovation Sciences
(2009).
\doiurl{10.6100/IR644335}
\end{botherref}
\endbibitem

\bibitem[\protect\citeauthoryear{{Van Houdt} et~al.}{2024}]{vanhoudt2024}
\begin{barticle}
\bauthor{\bsnm{{Van Houdt}}, \binits{G.}},
\bauthor{\bsnm{{de Leoni}}, \binits{M.}},
\bauthor{\bsnm{Martin}, \binits{N.}},
\bauthor{\bsnm{Depaire}, \binits{B.}}:
\batitle{An empirical evaluation of unsupervised event log abstraction techniques in process mining}.
\bjtitle{Information Systems}
\bvolume{121},
\bfpage{102320}
(\byear{2024})
\end{barticle}
\endbibitem

\bibitem[\protect\citeauthoryear{Basmer et~al.}{2024}]{basmer2024classification}
\begin{bchapter}
\bauthor{\bsnm{Basmer}, \binits{M.}},
\bauthor{\bsnm{Kabierski}, \binits{M.}},
\bauthor{\bsnm{Sahling}, \binits{K.}},
\bauthor{\bsnm{Patecka}, \binits{A.}},
\bauthor{\bsnm{Bala}, \binits{S.}},
\bauthor{\bsnm{Mendling}, \binits{J.}}:
\bctitle{A classification of data quality issues in object-centric event data}.
(\byear{2024})
\end{bchapter}
\endbibitem

\bibitem[\protect\citeauthoryear{Russell et~al.}{2006}]{russell2006workflow}
\begin{botherref}
\oauthor{\bsnm{Russell}, \binits{N.}},
\oauthor{\bsnm{Ter~Hofstede}, \binits{A.H.}},
\oauthor{\bsnm{Van Der~Aalst}, \binits{W.M.}},
\oauthor{\bsnm{Mulyar}, \binits{N.}}:
Workflow control-flow patterns: A revised view
(2006)
\end{botherref}
\endbibitem

\bibitem[\protect\citeauthoryear{Dijkman et~al.}{2011}]{dijkman2011similarity}
\begin{barticle}
\bauthor{\bsnm{Dijkman}, \binits{R.}},
\bauthor{\bsnm{Dumas}, \binits{M.}},
\bauthor{\bsnm{Van~Dongen}, \binits{B.}},
\bauthor{\bsnm{K{\"a}{\"a}rik}, \binits{R.}},
\bauthor{\bsnm{Mendling}, \binits{J.}}:
\batitle{Similarity of business process models: Metrics and evaluation}.
\bjtitle{Information Systems}
\bvolume{36}(\bissue{2}),
\bfpage{498}--\blpage{516}
(\byear{2011})
\end{barticle}
\endbibitem

\bibitem[\protect\citeauthoryear{Yujian and Bo}{2007}]{yujian2007normalized}
\begin{barticle}
\bauthor{\bsnm{Yujian}, \binits{L.}},
\bauthor{\bsnm{Bo}, \binits{L.}}:
\batitle{A normalized {Levenshtein} distance metric}.
\bjtitle{IEEE transactions on pattern analysis and machine intelligence}
\bvolume{29}(\bissue{6}),
\bfpage{1091}--\blpage{1095}
(\byear{2007})
\end{barticle}
\endbibitem

\bibitem[\protect\citeauthoryear{Gao et~al.}{2010}]{gao2010survey}
\begin{barticle}
\bauthor{\bsnm{Gao}, \binits{X.}},
\bauthor{\bsnm{Xiao}, \binits{B.}},
\bauthor{\bsnm{Tao}, \binits{D.}},
\bauthor{\bsnm{Li}, \binits{X.}}:
\batitle{A survey of graph edit distance}.
\bjtitle{Pattern Analysis and applications}
\bvolume{13},
\bfpage{113}--\blpage{129}
(\byear{2010})
\end{barticle}
\endbibitem

\bibitem[\protect\citeauthoryear{Sommers et~al.}{2022}]{sommers2022aligning}
\begin{bchapter}
\bauthor{\bsnm{Sommers}, \binits{D.}},
\bauthor{\bsnm{Sidorova}, \binits{N.}},
\bauthor{\bsnm{Dongen}, \binits{B.F.v.}}:
\bctitle{Aligning event logs to {Resource-Constrained} {Petri} nets}.
In: \bbtitle{International Conference on Applications and Theory of {Petri} Nets and Concurrency},
vol. \bseriesno{13288},
pp. \bfpage{325}--\blpage{345}.
\bpublisher{Springer},
\blocation{Cham}
(\byear{2022})
\end{bchapter}
\endbibitem

\bibitem[\protect\citeauthoryear{Sommers et~al.}{2024}]{sommers2024conformance}
\begin{bchapter}
\bauthor{\bsnm{Sommers}, \binits{D.}},
\bauthor{\bsnm{Sidorova}, \binits{N.}},
\bauthor{\bsnm{Dongen}, \binits{B.F.v.}}:
\bctitle{Conformance checking with model projections - rethinking log-model alignments for processes with interacting objects}.
In: \bbtitle{International Conference on Applications and Theory of {Petri} Nets and Concurrency},
vol. \bseriesno{14628}.
\bpublisher{Springer},
\blocation{Cham}
(\byear{2024})
\end{bchapter}
\endbibitem

\bibitem[\protect\citeauthoryear{van~der Werf et~al.}{2022}]{van2022data}
\begin{bchapter}
\bauthor{\bsnm{Werf}, \binits{J.M.E.}},
\bauthor{\bsnm{Rivkin}, \binits{A.}},
\bauthor{\bsnm{Polyvyanyy}, \binits{A.}},
\bauthor{\bsnm{Montali}, \binits{M.}}:
\bctitle{Data and process resonance: Identifier soundness for models of information systems}.
In: \bbtitle{International Conference on Applications and Theory of Petri Nets and Concurrency},
pp. \bfpage{369}--\blpage{392}.
\bpublisher{Springer}, \blocation{???}
(\byear{2022})
\end{bchapter}
\endbibitem

\bibitem[\protect\citeauthoryear{Ehrig and Padberg}{2003}]{ehrig2003graph}
\begin{bchapter}
\bauthor{\bsnm{Ehrig}, \binits{H.}},
\bauthor{\bsnm{Padberg}, \binits{J.}}:
\bctitle{Graph grammars and petri net transformations}.
In: \bbtitle{Advanced Course on Petri Nets},
pp. \bfpage{496}--\blpage{536}.
\bpublisher{Springer},
\blocation{Cham}
(\byear{2003})
\end{bchapter}
\endbibitem

\bibitem[\protect\citeauthoryear{Leemans et~al.}{2021}]{leemans2021stochastic}
\begin{barticle}
\bauthor{\bsnm{Leemans}, \binits{S.J.}},
\bauthor{\bsnm{Aalst}, \binits{W.M.}},
\bauthor{\bsnm{Brockhoff}, \binits{T.}},
\bauthor{\bsnm{Polyvyanyy}, \binits{A.}}:
\batitle{Stochastic process mining: Earth movers’ stochastic conformance}.
\bjtitle{Information Systems}
\bvolume{102},
\bfpage{101724}
(\byear{2021})
\end{barticle}
\endbibitem

\bibitem[\protect\citeauthoryear{Razouk and Phelps}{1983}]{razouk1983performance}
\begin{botherref}
\oauthor{\bsnm{Razouk}, \binits{R.R.}},
\oauthor{\bsnm{Phelps}, \binits{C.V.}}:
Performance analysis using timed {Petri} nets
(1983)
\end{botherref}
\endbibitem

\bibitem[\protect\citeauthoryear{Zuberek}{1980}]{zuberek1980timed}
\begin{bchapter}
\bauthor{\bsnm{Zuberek}, \binits{W.M.}}:
\bctitle{Timed {Petri} nets and preliminary performance evaluation}.
In: \bbtitle{Proceedings of the 7th Annual Symposium on Computer Architecture},
pp. \bfpage{88}--\blpage{96}
(\byear{1980})
\end{bchapter}
\endbibitem

\bibitem[\protect\citeauthoryear{Holliday and Vernon}{1987}]{holliday1987generalized}
\begin{botherref}
\oauthor{\bsnm{Holliday}, \binits{M.A.}},
\oauthor{\bsnm{Vernon}, \binits{M.K.}}:
A generalized timed {Petri} net model for performance analysis.
IEEE Transactions on Software Engineering
(12),
1297--1310
(1987)
\end{botherref}
\endbibitem

\bibitem[\protect\citeauthoryear{Sommers et~al.}{2023}]{sommers2023trident}
\begin{bchapter}
\bauthor{\bsnm{Sommers}, \binits{D.}},
\bauthor{\bsnm{Varadarajan}, \binits{D.V.}},
\bauthor{\bsnm{Sidorova}, \binits{N.}}:
\bctitle{Trident: Generating noisy synthetic processes with ground truth.}
In: \bbtitle{ICPM Doctoral Consortium/Demo}
(\byear{2023})
\end{bchapter}
\endbibitem

\bibitem[\protect\citeauthoryear{Peterson}{1981}]{peterson1981petri}
\begin{bbook}
\bauthor{\bsnm{Peterson}, \binits{J.L.}}:
\bbtitle{{Petri} Net Theory and the Modeling of Systems}.
\bpublisher{Prentice Hall PTR},
\blocation{Upper Saddle River}
(\byear{1981})
\end{bbook}
\endbibitem

\bibitem[\protect\citeauthoryear{Murata}{1989}]{murata1989petri}
\begin{barticle}
\bauthor{\bsnm{Murata}, \binits{T.}}:
\batitle{{Petri} nets: Properties, analysis and applications}.
\bjtitle{Proceedings of the IEEE}
\bvolume{77}(\bissue{4}),
\bfpage{541}--\blpage{580}
(\byear{1989})
\end{barticle}
\endbibitem

\bibitem[\protect\citeauthoryear{Rosa-Velardo and de~Frutos-Escrig}{2010}]{rosa2010decision}
\begin{botherref}
\oauthor{\bsnm{Rosa-Velardo}, \binits{F.}},
\oauthor{\bsnm{Frutos-Escrig}, \binits{D.}}:
Decision problems for {Petri} nets with names.
arXiv preprint arXiv:1011.3964
(2010)
\end{botherref}
\endbibitem

\bibitem[\protect\citeauthoryear{Sommers et~al.}{2023}]{sommers2023exact}
\begin{bchapter}
\bauthor{\bsnm{Sommers}, \binits{D.}},
\bauthor{\bsnm{Sidorova}, \binits{N.}},
\bauthor{\bsnm{Dongen}, \binits{B.}}:
\bctitle{Exact and approximated log alignments for processes with inter-case dependencies}.
In: \bbtitle{International Conference on Applications and Theory of Petri Nets and Concurrency},
pp. \bfpage{99}--\blpage{119}.
\bpublisher{Springer}, \blocation{???}
(\byear{2023})
\end{bchapter}
\endbibitem

\bibitem[\protect\citeauthoryear{Gianola et~al.}{2024}]{gianola2024object}
\begin{bchapter}
\bauthor{\bsnm{Gianola}, \binits{A.}},
\bauthor{\bsnm{Montali}, \binits{M.}},
\bauthor{\bsnm{Winkler}, \binits{S.}}:
\bctitle{Object-centric conformance alignments with synchronization}.
In: \bbtitle{International Conference on Advanced Information Systems Engineering},
pp. \bfpage{3}--\blpage{19}.
\bpublisher{Springer},
\blocation{Cham}
(\byear{2024})
\end{bchapter}
\endbibitem

\bibitem[\protect\citeauthoryear{Aalst and Berti}{2020}]{van2020discovering}
\begin{barticle}
\bauthor{\bsnm{Aalst}, \binits{W.M.v.}},
\bauthor{\bsnm{Berti}, \binits{A.}}:
\batitle{Discovering object-centric {Petri} nets}.
\bjtitle{Fundamenta informaticae}
\bvolume{175}(\bissue{1-4}),
\bfpage{1}--\blpage{40}
(\byear{2020})
\end{barticle}
\endbibitem

\bibitem[\protect\citeauthoryear{Fahland}{2019}]{fahland2019describing}
\begin{bchapter}
\bauthor{\bsnm{Fahland}, \binits{D.}}:
\bctitle{Describing behavior of processes with many-to-many interactions}.
In: \bbtitle{International Conference on Applications and Theory of {Petri} Nets and Concurrency},
pp. \bfpage{3}--\blpage{24}.
\bpublisher{Springer},
\blocation{Cham}
(\byear{2019})
\end{bchapter}
\endbibitem

\bibitem[\protect\citeauthoryear{van~der Aalst et~al.}{2004}]{DBLP:journals/tkde/AalstWM04}
\begin{barticle}
\bauthor{\bsnm{Aalst}, \binits{W.M.P.}},
\bauthor{\bsnm{Weijters}, \binits{T.}},
\bauthor{\bsnm{Maruster}, \binits{L.}}:
\batitle{Workflow mining: Discovering process models from event logs}.
\bjtitle{{IEEE} Trans. Knowl. Data Eng.}
\bvolume{16}(\bissue{9}),
\bfpage{1128}--\blpage{1142}
(\byear{2004})
\end{barticle}
\endbibitem

\bibitem[\protect\citeauthoryear{Leemans et~al.}{2013}]{DBLP:conf/bpm/LeemansFA13}
\begin{bchapter}
\bauthor{\bsnm{Leemans}, \binits{S.J.J.}},
\bauthor{\bsnm{Fahland}, \binits{D.}},
\bauthor{\bsnm{Aalst}, \binits{W.M.P.}}:
\bctitle{Discovering block-structured process models from event logs containing infrequent behaviour}.
In: \bbtitle{{BPM} 2013 Workshops}.
\bsertitle{LNBIP},
vol. \bseriesno{171},
pp. \bfpage{66}--\blpage{78}
(\byear{2013})
\end{bchapter}
\endbibitem

\bibitem[\protect\citeauthoryear{Carmona et~al.}{2018}]{carmona2018conformance}
\begin{barticle}
\bauthor{\bsnm{Carmona}, \binits{J.}},
\bauthor{\bsnm{Dongen}, \binits{B.}},
\bauthor{\bsnm{Solti}, \binits{A.}},
\bauthor{\bsnm{Weidlich}, \binits{M.}}:
\batitle{Conformance checking}.
\bjtitle{Switzerland: Springer}
\bvolume{56},
\bfpage{12}
(\byear{2018})
\end{barticle}
\endbibitem

\end{thebibliography}

\end{document}